\documentclass[graybox]{svmult}

\usepackage{type1cm}        
\usepackage{makeidx}         
\usepackage{graphicx}        
\usepackage{multicol}        
\usepackage[bottom]{footmisc}

\usepackage{newtxtext}       %
\usepackage{newtxmath}       

\usepackage{amsfonts,amsmath}

\usepackage{pdfsync,mathrsfs,fge}
\usepackage{mathdesign}

 \usepackage{theorem}

\usepackage{cancel}
\usepackage[normalem]{ulem}

\usepackage{graphics,graphicx,fancyhdr,color}
\usepackage{mathtools}

\usepackage{enumerate}

\newcommand{\bbZ}{{\mathbb Z}}

\newcommand{\bbR}{{\mathbb R}}

\newcommand{\bbT}{{\mathbb T}}
\newcommand{\eps}{\varepsilon}

\newcommand{\om}{\omega}

\newcommand{\bW}{{\bf W}}
\newcommand{\ga}{\gamma}
\newcommand{\la}{\lambda}
\newcommand{\al}{\alpha}
\newcommand{\bbE}{\mathbb E}
\newcommand{\farc}{\frac}

\theoremnumbering{arabic}
\theoremstyle{break}
\theoremsymbol{}
\theoremheaderfont{\normalfont\bfseries}
\theorembodyfont{\upshape}

\theoremstyle{nonumberbreak}
\newtheorem{remarknn}{Remark}

\makeindex             

\begin{document}

\title*{Thermal boundaries in kinetic and hydrodynamic limits}
\titlerunning{Thermal boundaries}
\author{Tomasz Komorowski and Stefano Olla}
\authorrunning{Komorowski and Olla} 
\institute{Tomasz Komorowski \at   Institute of Mathematics, Polish Academy of Sciences, ul. \'{S}niadeckich 8, 00-656 Warszawa \email{tkomorowski@impan.pl}
\and Stefano Olla \at CEREMADE, Universit\'e Paris Dauphine, PSL
Research University,   
{Institut Universitaire de France}, GSSI, L'Aquila, \email{olla@ceremade.dauphine.fr}}
%
%

\maketitle

\abstract*{We investigate how a thermal boundary, modelled by a Langevin dynamics, affect the macroscopic evolution of the energy at different space-time scales.
}

\abstract{We investigate how a thermal boundary, modelled by a Langevin dynamics, affect the macroscopic evolution of the energy at different space-time scales.
}

\section{Introduction}
\label{sec:intro}


Chains of an-harmonic oscillators are commonly {used   models in  non-equilibrium
statistical mechanics,  in particular to study macroscopic energy transport.}
To treat mathematically  non-linear dynamics is a very hard task,
even for a small non linear perturbation of the harmonic chain, see \cite{spohn06}.
In the purely harmonic chain the energy transport is ballistic, 
{see \cite{RLL}. Numerical evidence, see e.g. \cite{LLP}, shows that
non-linear perturbations can cause the transport {in a one-dimensional
system} to become diffusive, in case of optical chains
and superdiffusive for acoustic chains.
Replacing the non-linearity by a stochastic exchange of momenta
between neighboring particles makes the problem mathematically treatable
(see the review \cite{bbjko16} and the references therein).
This stochastic exchange can be modelled in various ways: e.g. for
each pair of the nearest neighbor particles the exchange of their momenta can occur independently at an exponential rate (which models their elastic collision). Otherwise, for each triple of consecutive particles, exchange of momenta can be performed in a continuous, diffusive fashion, so that its energy and momentum are preserved.
{The energy transport proven for such stochastic dynamics is qualitatively
similar to the one  expected  in the case of the non-linear
deterministic dynamics.} In particular, 
for a one-dimensional  acoustic chain it could be proved that the macroscopic thermal
energy density evolves according to a fractional heat equation
corresponding to the fractional laplacian
$(-\Delta)^{3/4}$, see \cite{jko15}.

In the recent years we have been interested in the macroscopic
effects of a heat bath in contact with the chain at a point.

In Section \ref{sec:micdyn} we consider  first a purely
harmonic chain in contact with a stochastic
Langevin thermostat at temperature $T$, see \eqref{eq:bas2} and \eqref{011210}.
In the absence of a thermostat,
its  dynamics is completely integrable
and  the energy of each frequency (the Fourier mode $k$) is conserved.
Rescaling space-time by the same parameter, the energy of mode $k$,
localized by Wigner distribution $W(t,y,k)$, evolves according to a
linear transport equation
\begin{equation*}
  \partial_t W + \bar \om'(k) \partial_y W = 0,
\end{equation*}
with velocity $\bar\omega'(k)= \omega'(k)/2\pi$,
where $\omega(k)$
is the dispersion relation of the harmonic chain.
We can interpret $W(t,y,k)$ as the energy density
of
\emph{phonons} of mode $k$ at time $t$ in the position $y$.
{The presence of a Langevin thermostat results
  in the emergence of a boundary (interface) condition at $y=0$ (\cite{kors18}):
 \begin{equation}\label{eq:bc}
\begin{split}
W(t,0^+,k)&=p_-(k)W(t,0^+, -k)+p_+(k)W(t,0^-,k)+{\frak g}(k)T,
\quad\hbox{ for $k>0$}\\
&\\
W(t,0^-, k)&=p_-(k)W(t,0^-,-k) + p_+(k)W(t,0^+, k) +{\frak
  g}(k)T,\quad \hbox{ for $k<0$.}
\end{split}
\end{equation}
}
The coefficients appearing in the boundary condition correspond to 
probabilities of the  phonon
transmission $p_+(k)$, reflection $p_-(k)$ and absorption
$\fgeeszett(k)$.  These parameters are non-negative and  satisfy
$p_+(k) +  p_-(k) + \fgeeszett(k) = 1$.
In addition, $T \fgeeszett(k)$ is the intensity of the phonon creations. 
The transmission, reflection   and absorption  parameters depend in a quite
complicated way on the dispersion relation $\omega(\cdot)$ and the
strength of the thermostat $\ga>0$ (cf. \eqref{nu},  \eqref{eq:9},  \eqref{eq:23}).
Some of their properties   and an explicit calculation for the nearest-neighbor
interactions are presented in Appendix A (see Section \ref{sec:prop-scatt-coeff},
the results contained  there are original and
 are not part of \cite{kors18}).

It is somewhat surprising  that an incident \emph{phonon} of mode $k$
after scattering, if not absorbed,
can produce only an identical transmitted phonon, or a reflected phonon
of mode $-k$ (at least for a unimodal dispersion relation).
This stands in contrast with what takes place at the microscopic scale. Then, an incident
wave of frequency $k$ scatters and produces waves 
of all possible frequencies. 
{ In the macroscopic limit, all frequencies produced
  by the scattering on the thermostat,
except  those corresponding to  $\pm k$, are damped by oscillations.}

In \cite{ko19} we have considered the same problem after adding
the bulk noise that conserves energy and momentum, see Section \ref{sec:harmonic-chain-with}.
The noise is properly rescaled in such a way that  finite total amount
of momentum is exchanged { locally}
in the macroscopic  unit time  (in
analogy to a kinetic limit).
{The effect of the bulk noise is to add a macroscopic scattering
term to the transport equation : 
\begin{equation}\label{eq:B}
  \partial_t W + \bar \om'(k) \partial_y W =
  \gamma_0\int_{\bbT} R(k,k')\left(W(k') - W(k)\right) dk',
\end{equation}
i.e. a phonon of mode $k$ changes to the one of mode $k'$, with intensity $\gamma_0 R(k,k')$,
given by \eqref{eq:R}.
The case  when no the heat bath is present, has been studied in \cite{bos09}.
In \cite{ko19} we prove that the heat bath adds to \eqref{eq:B} the same boundary condition
\eqref{eq:bc}, see Section \ref{sec:harmonic-chain-with}.}

Since in \eqref{eq:B} the energies of the different modes are mixed up by the bulk scattering,
we can further rescale space-time in this equation
in order to obtain an autonomous equation for the
evolution of the total energy. Without the thermal bath this case has been studied in
\cite{jko09} and the following results have been obtained:
\begin{itemize}
\item For \emph{optical chains} the velocity of the phonon behaves like $\bar\omega'(k) \sim k$
for small $k$, while for the {total} scattering rate $R(k) = \int R(k,k') dk' \sim k^2$. This means that
phonons of low frequency rarely scatter, but move very slowly so they have the time to
diffuse under an appropriate (diffusive) scaling. The phonons
corresponding to other modes also behave diffusively at the respective scales. Consequently,
in the optical chain, under diffusive space-time scaling, all modes homogenize equally,
contributing to the macroscopic evolution of the
energy $e(t,y)$, i.e. $W(t/\delta^2, y/\delta, k) \mathop{\longrightarrow}_{\delta\to 0} e(t,y)$,
that follows the linear heat equation
$(\partial_t-D\partial_y^2)e(t,y)=0$ with an explicitly given $D>0$,
see \eqref{eq:D}.
\item In \emph{acoustic chains},
the bulk scattering rate is the same, but  $\bar\omega'(k) \sim O(1)$ for small $k$.
Consequently low frequency phonons scatter rarely but they still move with velocities of order 1,
and the resulting macroscopic limit is superdiffusive.
In particular,
in this case the low frequency modes
are responsible for the macroscopic transport of the energy.
{The respective superdiffusive space-time scaling limit
  $W(t/\delta^{3/2}, y/\delta, k) \mathop{\longrightarrow}_{\delta\to 0} e(t,y)$,
  described by the solution of  a fractional heat equation
 $(\partial_t-\hat c|\partial_y^2|^{3/4})e(t,y)=0$, with $\hat c>0$
  given by \eqref{hatc}.}
\end{itemize}

The thermal bath adds a boundary condition at $y=0$ to the 
diffusive, or superdiffusive equations described above. More precisely the situation is as follows. 
\begin{itemize}
\item In the optical chain we obtain a Dirichlet boundary condition
  $e(t,0) = T$ for the respective heat equation (see \eqref{eq:26}):
  phonons {trajectories} behave like Brownian motions, and since they can cross
  the boundary infinitely many times, they are absorbed almost
  surely. In effect,
  there is no energy {\emph{transfer}} through the boundary at the macroscopic scale.
  This is proven in \cite{bko19} using analytic techniques, see
  Section \ref{sec:optic-chain:-diff}.

\item In the acoustic chain, the long wave phonons, are responsible for the macroscopic
  energy transport. Their trajectories behave in the limit like
  superdiffusive symmetric, $3/2$--stable Levy processes: they can jump over the boundary
  on the macroscopic scale and there is a positive probability of survival, i.e.
  of energy macroscopic transmission across the thermal boundary.
  Since the absorption probability $\fgeeszett(k)$ remains strictly positive as $k\to 0$
  (as we prove here in appendix A, at least for the nearest neighbor acoustic chain,
  see \eqref{eq:23a}),  the thermal boundary affects the transport. The macroscopic
  energy evolution is given by a fractional heat equation with boundary defined by
  \eqref{eq:31} and \eqref{apr204}. 
 This is proven in \cite{kor19} using probabilistic techniques, see
 Section \ref{sec:acoust-chain:-superd}.
\end{itemize}

In Section 2 we review the results of  \cite{kors18} for the harmonic chain with the
thermostat attached at a point. Since the calculations for the transmission and reflection
scattering of  \cite{kors18} are quite complex, {we present their
somewhat simplified version (conveying nevertheless their gist)} in Appendix B (Section
\ref{app:2}). We hope
that the outline  would help the reader to understand how the
macroscopic scattering  emerges.

In Section 3 we review the results of \cite{ko19} in the presence of the conservative
bulk noise. Section 4 contains the review of the diffusive and superdiffusive limits proven in
\cite{bko19} and  \cite{kor19}. In Section 5 we mention some open problems, in particular
the question of the direct hydrodynamic limit, without passing through the kinetic limit, in the spirit of \cite{jko15}.

Appendix A (Section  \ref{sec:prop-scatt-coeff})  contains some
original results that are not present in the discussed articles, concerning the properties
of the scattering coefficients and their behaviour for $k\to 0$.

\section{Notation}
  
Given $a>0$ by $\bbT_a$ we denote the torus of size $a>0$, i.e. the interval $[-a/2,a/2]$
with identified endpoints. When $a=1$ we shall write $\bbT:=\bbT_1$
for the unit torus. Let $\bbT_\pm:=[k\in\bbT:\,1/2>\pm k>0]$. We also
let $\bbR_+:=(0,+\infty)$, $\bbR_*:=\bbR\setminus\{0\}$ and $\bbT_*:=\bbT\setminus\{0\}$.

{By $\ell^p(\bbZ)$, $L^p(\bbT)$, where $p\ge1$, we denote the spaces of all complex
 valued sequences $(f_x)_{x\in\bbZ}$ and functions $f:\bbT\to\mathbb
 C$ that are summable with $p$-th power, respectively.}  The Fourier transform of  $(f_x)_{x\in\bbZ}\in \ell^2(\bbZ)$
  and the inverse Fourier transform of $\hat f\in L^2(\bbT)$ are given by
  \begin{equation}
  \label{fourier}
  \hat f(k)=\sum_{x\in\bbZ} f_x \exp\{-2\pi ixk\}, ~~
f_x=\int_{\bbT} \hat f(k) \exp\{2\pi ixk\} dk, \quad x\in \bbZ,~~k\in\bbT.
\end{equation} 
We use the notation 
\[
(f\star g)_y=\sum_{y'\in\bbZ}f_{y-y'}g_{y'}
\]
for the convolution of two sequences  $(f_x)_{x\in\bbZ},
(g_x)_{x\in\bbZ}\in \ell^2(\bbZ)$ that belong to appropriate spaces
$\ell^p(\bbZ)$. {In most cases we shall assume that one of the sequences
rapidly decays, while the other belongs to  $\ell^2(\bbZ)$.}

For a function $G:\bbR\times\bbT\to\mathbb C$ that is either $L^1$, or $L^2$-summable, we denote by 
$\hat
G:\bbR\times\bbT\to\mathbb C$ its Fourier transform, in the first variable, defined as 
\begin{equation}
    \hat G(\eta,k):=\int_{\bbR}e^{-2\pi i \eta x}G(x,k)dx,\quad (\eta,k)\in\bbR\times \bbT.
    \label{eq:2}
\end{equation}
Denote by $C_0(\bbR\times\bbT)$ the class of functions $G$ that are
continuous and satisfy $\lim_{|y|\to+\infty}\sup_{k\in\bbT}|G(y,k)|=0$.

\section{Harmonic chain in contact with a Langevin thermostat}
\label{sec:micdyn}

We consider the evolution of an infinite particle system governed by the Hamiltonian 
\begin{equation}
\label{011210}
{\cal H}({\frak p},{\frak q}):=\frac12\sum_{y\in\bbZ}{\frak p}_y^2+\frac{1}{2}\sum_{y,y'\in\bbZ}\alpha_{y-y'}{\frak q}_y{\frak q}_{y'}.
\end{equation}
Here, the  particle label is $y\in\bbZ$, $({\frak q}_y,{\frak p}_y)$ is the position and momentum of the $y$'s particle, respectively, and $({\frak q},{\frak p})=\{({\frak q}_y,{\frak p}_y),\,y\in\bbZ\}$ denotes the entire configuration.
The coupling coefficients $\alpha_{y}$ are assumed
to have exponential decay and chosen positive definite such that the energy
is positive. 
We  couple the particle with label $0$ to a Langevin thermostat at temperature~$T$.
Then the evolution equation then writes as the stochastic differential equations:
\begin{equation}
  \begin{split}
&\dot{\frak q}_y(t)={\frak p}_y(t),
\label{eq:bas2}\\
& d{\frak p}_y(t) = -(\alpha\star {\frak q}(t))_ydt
+\big(-\ga {\frak p}_0(t)dt+\sqrt{2\ga T}dw(t)\big) \delta_{0,y},\quad y\in\bbZ.
\end{split}
\end{equation}
Here, $\{w(t),\,t\ge0\}$ is a
standard Wiener process, 
while $\gamma>0$ is a coupling parameter with the thermostat. 

\subsubsection* {Assumptions on the dispersion relation and its basic properties}

We assume  (cf  \cite{bos09}) that the coupling constants
$(\al_x)_{x\in\bbZ}$ satisfy the following:
 \begin{enumerate}
 \item[a1)] they are real valued and there exists $C>0$ such that $|\alpha_x|\le Ce^{-|x|/C}$ for all $x\in \bbZ$,
  \item[a2)] $\hat\alpha(k)=\sum_{x\in\bbZ}\al_xe^{-2\pi i kx}$ is also real valued 
and  $\hat\alpha(k)>0$ for $k\not=0$ and in case $\hat \alpha(0)=0$ we  have  $\hat\alpha''(0)>0$.
 \end{enumerate}
   The above conditions imply that both functions $x\mapsto\alpha_x$
   and $k\mapsto\hat\alpha(k)$ are   even. In addition, $\hat\alpha\in
   C^{\infty}(\bbT)$ and in case $\hat\alpha(0)=0$ we have
   $\hat\alpha(k)=k^2\phi(k^2)$ for some strictly positive  $\phi\in
   C^{\infty}(\bbT)$.  The  dispersion relation  $\om:\bbT\to
   \bar\bbR_+$, given by
 \begin{equation}\label{mar2602}
 \om(k):=\sqrt{\hat \alpha (k)},\quad k\in\bbT.
\end{equation}
is obviously also even.
Throughout the paper it is assumed to be unimodal, i.e. increasing on
$\bar\bbT_+$ and then, in consequence, decreasing on
$\bar\bbT_-$. Its 
unique minimum and maximum are attained at $k=0$, $k=1/2$,
respectively. They
are denoted by~$\om_{\rm min}\ge 0$ and $\om_{\rm max}$,
correspondingly. Denote the two branches of its inverse
by~$\om_\pm:[\om_{\rm min},\om_{\rm max}]\to\bar \bbT_\pm$.

In order to avoid technical problems with the definition of the dynamics, we assume
that the initial conditions are random but with
finite energy: ${\cal H}({\frak p},{\frak q}) < \infty$. This property will be conserved in time.
For such configurations we can define the complex wave function
\begin{equation}
\label{011307}
\psi_y(t) := (\tilde{\om} \star{\frak q}(t))_ y + i{\frak p_y}(t)
\end{equation}
where $\big(\tilde \om_y\big)_{y\in\bbZ}$ is the inverse
 Fourier transform of the  dispersion relation. 
We have ${\cal H}({\frak p}(t),{\frak q}(t)) = \sum_y |\psi_y(t)|^2$. 

The Fourier transform of the wave function is given by 
\begin{equation}
\label{011307a}
\hat\psi(t,k) := \om(k) \hat {\frak q}(t,k) + i\hat{\frak
  p}(t,k),\quad k\in\bbT,
\end{equation}
so that
$$
\hat{\frak p}\left(t,k\right)=\frac{1}{2i}[\hat\psi(t,k)-\hat\psi^*(t,-k)],
~~{\frak p}_0(t)=\int_{\bbT} {\rm Im}\,\hat\psi(t,k) dk.
$$
Using \eqref{eq:bas2}, it is easy to verify that the wave function evolves according to  
\begin{equation}
\begin{split}
 \label{basic:sde:2aa}
 d\hat\psi(t,k) &= \big(-i\om(k)\hat\psi(t,k)
 - i{\ga} {\frak p}_0(t) \big) dt
 +i\sqrt{2\ga T}dw(t).
\end{split}
\end{equation}

Introducing a (small) parameter $\eps\in(0,1)$,
we wish to study the behaviour of the distribution of the energy
at a large space-time scale, i.e. for the wave function $\psi_{[x/\eps]}(t/\eps)$,
$(t,x)\in \Bbb R_+\times\Bbb R$, {when $\eps\to0$. In this scaling
  limit we would like to maintain 
each particle contribution to the total energy  to be of order $O(1)$,
on the average, and therefore
keep the total energy of the chain to be order $\eps^{-1}$.}
For this reason,  we choose random initial data that is distributed by 
probability measures $\mu_\eps$  defined on the phase space $({\frak
  p},{\frak q})$, in such a way that
\begin{equation}
  \label{eq:3}
  \sup_{\eps\in(0,1)} \eps \langle{\cal H}({\frak p},{\frak q}) \rangle_{\mu_\eps}
  = \sup_{\eps\in(0,1)}\sum_{y\in\bbZ}\eps\langle|\psi_y|^2\rangle_{\mu_\eps}
  =\sup_{\eps\in(0,1)}\eps\langle \|\hat \psi\|^2_{L^2(\bbT)}\rangle_{\mu_\eps}
<\infty.
\end{equation}
The symbol $\langle\cdot\rangle_{\mu_\eps}$ denotes, as usual, the
average with respect to measure $\mu_\eps$.
To simplify our calculations we will also assume that 
\begin{equation}
\label{null}
\langle\hat\psi(k)\hat\psi(\ell) \rangle_{\mu_\eps}=0,\quad k,\ell\in\bbT. 
\end{equation}
This condition is easily satisfied by local Gibbs measures like
\begin{equation}
  \label{eq:LG}
  \prod_{y\in\Bbb Z} \frac{e^{-\beta^\eps_y |\psi_y|^2/2}}{Z_{\beta^\eps_y}} d \psi_y
\end{equation}
for a proper choice of temperature profiles $(\beta^\eps_y)^{-1}>0$,
decaying fast enough to 0, as $|y|\to+\infty$. {Here $Z_{\beta^\eps_y}$ is the normalizing constant.}

\subsubsection*{Wigner distributions}

Wigner distributions provide an effective tool to localize in space
energy per frequency, separating microscopic from macroscopic scale.
The {(averaged)} Wigner distribution (or Wigner transform) is
defined by its action on a test function 
$G \in {\cal S}(\bbR\times\bbT)$~as
\begin{equation}
\label{wigner}
\langle G,W^{(\eps)}(t)\rangle:=\frac{\eps}{2}
\sum_{y,y'\in\bbZ}\int_{\bbT}e^{2\pi ik(y'-y)}\bbE_\eps\left[\psi_y \left(\frac{t}{\eps}\right)
\left(\psi_{y'}\right)^*\left(\frac{t}{\eps}\right)\right]
G^*\Big(\eps\frac{y+y'}{2},k\Big)dk. 
\end{equation}

The Fourier transform of the Wigner distribution, {or the
  Fourier-Wigner function} is defined as
\begin{equation}
  \label{eq:20}
  \widehat{ W}_\varepsilon(t,\eta,k) \ :=\ 
  \frac{ \varepsilon}{2} \bbE_\eps
  \left[ \hat\psi^*\left( \frac{t}{\eps} , k- \frac{\varepsilon\eta}2\right) 
    \hat\psi\left( \frac{t}{\eps}  , k +  \frac{\varepsilon\eta}2\right)\right],
  \quad (t,\eta,k)\in [0,\infty)\times\bbT_{2/\eps}\times\bbT,
\end{equation}
so that
\begin{equation}
\label{wigner1}
\langle G,W^{(\eps)}(t)\rangle=\int_{\bbT\times\bbR}
\widehat{ W}_\varepsilon(t,\eta,k)\hat G^*(\eta,k)d\eta dk,
\quad G \in {\cal S}(\bbR\times\bbT).
\end{equation}
Taking $G(x,k) := G(x)$ in  \eqref{wigner} we obtain
\begin{equation}
  \label{eq:1}
  \langle G,W^{(\eps)}(t)\rangle= \frac{\eps}{2}
\sum_{y\in\bbZ}\bbE_\eps\left[\left|\psi_y \left(\frac{t}{\eps}\right)\right|^2 \right]
G(\eps y). 
\end{equation}

{
In what follows we assume that the initial data, after averaging,
leads to a sufficiently
fast decaying (in $\eta$) Fourier-Wigner function. More precisely, we
suppose that 
there exist $C,\kappa>0$ such that 
\begin{equation}
\label{011812aa}
|\widehat W_\eps(0,\eta,k)|
\le 
 \frac{C}{(1+\eta^2)^{3/2+\kappa}},
\quad (\eta,k)\in\bbT_{2/\eps}\times \bbT, \,\eps\in(0,1).
\end{equation} 
In addition, we assume that there exists a distribution $W_0\in {\cal
  S}'(\bbR\times\bbT)$ such that
for any $G \in {\cal S}(\bbR\times\bbT)$
\begin{equation}
\label{010709-20}
\lim_{\eps\to0+}\langle G,W^{(\eps)}(0)\rangle =\langle G, W_0\rangle.
\end{equation}
Note that, thanks to \eqref{011812aa}, distribution $ W_0$ is in fact a
function
that belongs to $C_0(\bbR\times\bbT)\cap L^2 (\bbR\times\bbT)$.
}

\subsection{The thermostat free case: $\gamma = 0$}
\label{sec:free-case:-gamma}

If the thermostat is not present ($\gamma = 0$), the equation of motion
\eqref{basic:sde:2aa} can be explicitely solved and the soution is
$\hat\psi(t,k) = \hat\psi(k) e^{-i\om(k) t}$.
Defining
\begin{equation}
\label{053110}
\delta_{\eps}\om(k,\eta):=\frac{1}{\eps}\left[\om\left(k+\frac{\eps
      \eta}{2}\right)-\om\left(k-\frac{\eps \eta}{2}\right)\right],
\end{equation}
we can compute explicitly the Wigner transform:
\begin{equation}
  \label{eq:5}
  \widehat{ W}_\varepsilon(t,\eta,k) \ =\
  e^{-i\delta_{\eps}\om(k,\eta)t/\eps}
  \widehat{ W}_\varepsilon(0,\eta,k)
  \mathop{\longrightarrow}_{\varepsilon\to 0} e^{-i\omega'(k)\eta t} \widehat{ W}_0(\eta,k),
\end{equation}
assuming the corresponding convergence at initial time, see \eqref{010709-20}.
The inverse Fourier transform
gives
\begin{equation}
  \label{eq:18}
    W (t,y ,k) \ =\ W_0(y - \bar \omega'(k) t ,k),
  \end{equation}
  where $\bar \omega'(k) := \om'(k)/(2\pi)$,
i.e. it solves the simple
linear transport equation
\begin{equation}
  \label{eq:6}
  \partial_t W(t,y,k) + \bar \om'(k) \partial_y W(t,y,k) = 0,\quad W(0,y,k)=W_0(y,k). 
\end{equation}
We can view this equation as the evolution of the density in
independent particles {(phonons)},
labelled by the frequency mode $k\in\bbT$, and
moving with velocity $\bar \omega'(k)$.

\subsection{The evolution with the Langevin thermostat: $\gamma>0$}
\label{sec:evol-with-lang}

We use the mild formulation of   \eqref{basic:sde:2aa}: 
\begin{equation}
  \label{eq:sol1}
  \begin{split}
    \hat\psi(t,k) = e^{-i\omega(k) t} \hat\psi(0,k) -
    i\gamma \int_0^t e^{-i\omega(k) (t-s)} {\frak p}_0(s) ds
    + i \sqrt{2\ga T} \int_0^t e^{-i\omega(k) (t-s)} dw(t).
  \end{split}
\end{equation}
Integrating both sides in the $k$-variable and taking the imaginary part in both sides, we obtain a closed equation 
for ${\frak p}_0(t)$:
\begin{equation}
  \label{eq:p0}
  \begin{split}
    {\frak p}_0(t) &= {\frak p}^0_0(t) - \gamma  \int_0^t J(t-s) {\frak p}_0(s) ds +  
    \sqrt{2\ga T} \int_0^t J(t-s) dw(s),
  \end{split}
\end{equation}
where \begin{equation}
  \label{eq:bessel0}
  J(t) = \int_{\bbT}\cos\left(\omega(k) t\right) dk,
\end{equation}
and 
\begin{equation}
  \label{eq:1}
   {\frak p}^0_0(t) = \int_{\bbT} {\rm Im}\left(\hat\psi(0,k) e^{-i\omega(k) t}\right) dk,
\end{equation}
is the  momentum at   $y=0$   for the free evolution 
with $\gamma = 0$ (without the thermostat).

Taking the Laplace transform
$$
\tilde {\frak p}_0 (\lambda) =\int_0^{+\infty}e^{-\la t} {\frak p}_0(t) dt,\quad {\rm Re}\,\la>0,
$$
 in \eqref{eq:p0} we obtain
 \begin{equation}
  \label{eq:3}
  \tilde {\frak p}_0 (\lambda) =  \tilde g(\lambda) \tilde   {\frak p}^0_0(\lambda) + 
    \sqrt{2\ga T} \tilde g(\lambda) \tilde J(\lambda) \tilde w(\lambda).
\end{equation}
Here, $\tilde g(\lambda)$ is given by
\begin{equation}
\label{tg}
\tilde g(\lambda) := ( 1 + \gamma \tilde J(\lambda))^{-1}.
\end{equation}
and
\begin{equation}
  \label{eq:2}
\tilde J(\la):=\int_0^{\infty}e^{-\la t}J(t)dt= \int_{\bbT}  \frac{\lambda}{\lambda^2 + \omega^2(k)} dk,\quad {\rm Re}\,\la>0.
\end{equation}
{We will show below} 
that $\tilde g(\lambda)$ is the Laplace transform
of a signed locally finite measure~$g(d\tau)$. Then, the term
$(\lambda + i \omega(k))^{-1} \tilde g(\lambda) \tilde{\frak p}_0^0(\lambda)$,
that appears in (\ref{eq:6}),
is the Laplace transform~of the convolution
\begin{equation}
  \label{eq:11}
  \int_0^t \phi(t-s,k)  {\frak p}_0^0(s) ds,
\end{equation}
where 
\begin{equation}
  \label{eq:12}
   \phi(t,k)= \int_0^{t}e^{-i\omega(k)(t-\tau)} g(d\tau) .
 \end{equation}
Next, taking the Laplace transform of both sides of  \eqref{eq:sol1} and using \eqref{eq:3},  we arrive at an explicit formula for the Fourier-Laplace 
transform of $\psi_y(t)$:
\begin{equation}
  \label{eq:6}
  \begin{split}
    \tilde \psi(\lambda,k) &= 
\frac{\hat\psi(0,k) - i\gamma \tilde{\frak p}_0(\lambda) + i\sqrt{2\gamma T} \tilde w(\lambda)}
{\lambda + i\omega(k)} \\
& 
= \frac{\hat\psi(0,k) - i\gamma \tilde g(\lambda) (\tilde{\frak p}_0^0(\lambda)  
+ \sqrt{2\gamma T}  \tilde J(\lambda)\tilde w(\lambda)) + i  \sqrt{2\ga T} \tilde w(\lambda)}
{\lambda + i\omega(k)}  \\
& = \frac{\hat\psi(0,k) - i\gamma \tilde g(\lambda) \tilde{\frak p}_0^0(\lambda)  
+  i \tilde g(\lambda) \sqrt{2\gamma T} \tilde w(\lambda)}
{\lambda + i\omega(k)} .
  \end{split}
\end{equation}
The Laplace inversion of  \eqref{eq:6} yields an explicit expression for
$\hat\psi(t,k)$: 
\begin{equation}
  \label{eq:10}
  \begin{split}
    \hat\psi(t,k) =& e^{-i\omega(k) t} \hat\psi(0,k) - 
     i \gamma  \int_0^t \phi(t-s,k) {\frak p}_0^0(s)\; ds \\
   &+ i \sqrt{2\gamma T} \int_0^t \phi(t-s,k) \; dw(s).
  \end{split}
\end{equation}

\subsection{Phonon creation by the heat bath}
\label{sec:phonon-creation-heat}

Since the contribution to the energy given by the thermal term
and the initial energy are completely separate,
we can assume first that $\widehat W_0 = 0$.
In this case  $\hat\psi(0,k) = 0$
and \eqref{eq:10} reduces to a stochastic convolution:
\begin{equation}
  \label{eq:10T}
  \begin{split}
    \hat\psi(t,k) = i \sqrt{2\gamma T} \int_0^t \phi(t-s,k) \; dw(s).
  \end{split}
\end{equation}
To shorten the notation, denote 
\[
\tilde\phi(t, k) = \int_0^{t} e^{i\omega(k)\tau} g(d\tau) = e^{i\omega(k)t}\phi(t,k),
\]
We can compute directly the Fourier-Wigner function
\begin{equation*}
  \widehat W_\eps(t,\eta,k) = \gamma T \int_0^t 
  e^{-i \delta_{\eps}\om(k,\eta) s} \tilde\phi\left(s/\eps, k+\frac{\eps\eta}2\right)
  \tilde\phi^*\left(s/\eps, k-\frac{\eps\eta}2\right) ds.
\end{equation*}
Taking its Laplace transform we obtain
\begin{equation}
  \label{eq:24}
  \begin{split}
    \widehat{w}_\eps(\la,\eta,k) &= \gamma T \int_0^\infty dt
    e^{-\lambda t} \int_0^t ds e^{-i \delta_\eps\omega(k,\eta) s}
    \tilde\phi\left(\eps^{-1} s, k+\frac{\eps\eta}2\right) \tilde\phi^*\left(\eps^{-1} s, k-\frac{\eps\eta}2\right) \\
    &= \frac{\gamma T}{\lambda}  \int_0^\infty ds e^{-(\lambda +i \delta_\eps\omega(k,\eta)) s} 
    \tilde\phi\left(\eps^{-1} s, k+\frac{\eps\eta}2\right)
    \tilde\phi^*\left(\eps^{-1} s, k-\frac{\eps\eta}2\right). 
  \end{split}
\end{equation}
Using the inverse Laplace formula for the product of functions 
we obtain, for any $c>0$, 
\begin{equation}
  \label{eq:16}
  \begin{split}
  &  \widehat{w}_\eps(\la,\eta,k)  = \frac{\gamma T}{\lambda} \frac
    1{2\pi i}\lim_{\ell\to\infty} \int_{c-i\ell}^{c+i\ell} \big\{\sigma\left(\lambda + i \delta_\eps\omega(k,\eta) - \sigma\right)\big\}^{-1} 
    \\
&\times {\tilde g\left(\eps \sigma - i\omega(k+\frac{\eps\eta}{2}) \right) 
      \tilde g^*\left(\eps(\lambda + i \delta_\eps\omega(k,\eta) - \sigma) - i\omega(k-\frac{\eps\eta}{2}) \right)}
    \ d\sigma {.}
  \end{split}
\end{equation}
Since $\tilde g$ is bounded and $\text{Re} \lambda >0$,
we can take the limit as $\eps \to 0$, obtaining
\begin{equation}
  \label{eq:17}
  \widehat{w}(\la,\eta,k)  =
  \frac{\gamma T |\nu(k)|^2}{\lambda\left(\lambda + i \omega'(k) \eta\right)},
\end{equation}
where
\begin{equation}
\label{nu}
\nu(k) :=\lim_{\eps\to0}\tilde g(\eps-i\om(k)).
\end{equation}
The limit in \eqref{nu} is well defined everywhere, see Section \ref{sec:prop-scatt-coeff} for details.

The inverse Laplace transform of \eqref{eq:17} gives
\begin{equation}
  \label{eq:4}
  \widehat{W}(t,\eta,k)  = \frac{1-e^{-i\omega'(k)\eta t}}{i\omega'(k) \eta} \gamma T |\nu(k)|^2.
\end{equation}
Performing the inverse Fourier transform, according to \eqref{eq:2},
we obtain
\begin{equation}
  \label{eq:8}
  W(t,y,k) = T \fgeeszett(k) 1_{[[0, \bar\omega'(k)t]]}(y)
\end{equation}
where $\bar\omega(k) = \omega(k)/2\pi$, 
\begin{equation}
  \label{eq:9}
 \fgeeszett(k) :=  \frac{\gamma |\nu(k)|^2}{|\bar\omega'(k)|}
\end{equation} 
and 
\begin{equation*}
[[0, a]] := \begin{cases}
[0, a], &\text{if $a>0$}\\
[a, 0], &\text{if $a<0$.}
\end{cases}
\end{equation*}
We can interpret \eqref{eq:8} as the
energy density of $k$-phonons that are created at the interface $y=0$ by the heat
bath with intensity $T \fgeeszett(k)$
and then move with velocity $\bar\omega'(k)$.
The Wigner function $W(t,y,k)$ can be viewed as a formal solution of  
\begin{equation}
  \label{eq:7}
  \partial_t W(t,y,k) + \bar\omega'(k) \partial_y  W(t,y,k) =
  |\bar\omega'(k)| T \fgeeszett(k)\delta(y), \qquad W(0,y,k) = 0. 
\end{equation}


\subsection{Phonon scattering and absortion by the heat bath}
\label{sec:phon-scatt-therm}

The scattering of incoming waves can be studied at temperature $T=0$,
by looking at the deterministic equation
 \begin{equation}
  \label{eq:10-det}
  \begin{split}
    \hat\psi(t,k) 
   = e^{-i\omega(k) t} \hat\psi(0,k) - 
     i \gamma  \int_0^t \phi(t-s,k) {\frak p}_0^0(s)\; ds,
  \end{split}
\end{equation}
Proceeding 
along the lines of the calculation of the previous section we obtain
\begin{equation}
  \label{eq:5-det0}
  \begin{split}
  &\widehat{ W}_\varepsilon(t,\eta,k) \ = \
  \widehat{
    W}_\varepsilon^0(t,\eta,k) +\widehat{
    W}_\varepsilon^1(t,\eta,k) +\widehat{ W}_\varepsilon^2(t,\eta,k),
\end{split}
\end{equation}
where $\widehat{ W}_\varepsilon(t,\eta,k)$ is given by \eqref{eq:20},
\begin{equation}
\label{eq:5-det}
  \begin{split} 
&\widehat{W}_\varepsilon^0(t,\eta,k):= e^{-i\delta_{\eps}\om(k,\eta)t/\eps}
  \widehat{ W}_\varepsilon(0,\eta,k),\\
  &\widehat{
    W}_\varepsilon^1(t,\eta,k):= -i\frac{\eps\gamma}{2}
  \int_0^{t/\eps} \Bigg\{ \bbE_\eps\Big[\hat\psi\left(0, k-\frac{\eps\eta}{2}\right)^*
   {\frak p}_0^0(s)\Big]
  e^{i\omega(k-\frac{\eps\eta}{2}) \frac{t}{\eps}}
  \phi\left( \frac{t}{\eps}-s, k+\frac{\eps\eta}{2}\right)\\
     & \qquad \qquad- \bbE_\eps\Big[ \hat\psi\left(0, k+\frac{\eps\eta}{2}\right){\frak p}_0^0(s)\Big]
      e^{-i\omega(k+\frac{\eps\eta}{2}) \frac{t}{\eps}}
     \phi\left( \frac{t}{\eps}-s, k-\frac{\eps\eta}{2}\right)^*\Bigg\}\; ds ,\\
      &\widehat{W}_\varepsilon^2(t,\eta,k):= \frac{\eps\gamma^2}{2}
      \int_{0}^{t/\eps} ds_1   \int_{0}^{t/\eps} ds_2 
      \bbE_\eps  \left[{\frak p}_0^0(s_1) {\frak p}_0^0(s_2)\right]\\
    & \qquad \qquad\times\phi\left(\frac{t}{\eps}-s_1,k-\frac{\eps\eta}{2}\right)^*
     \phi\left(\frac{t}{\eps}-s_2,k-\frac{\eps\eta}{2}\right).
\end{split}
\end{equation}

The limit behavior of  $\widehat{W}_\varepsilon^0(t,\eta,k)$
is already described by \eqref{eq:5}.
The calculations for the other two terms $\widehat{W}_\varepsilon^1$
and $\widehat{W}_\varepsilon^2$ are more involved. Their
outline is presented in Section \ref{app:2} below. We have
\begin{equation}
  \label{eq:13b}
\lim_{\eps\to0+} \widehat{
    W}_\varepsilon^1(t,\eta,k)   =-  \gamma \text{Re} \nu(k) e^{-i\omega'(k)t} \int_{\bbR}
  \frac{1-e^{-i\omega'(k)(\eta'-\eta) t}}{i\omega'(k) (\eta'-\eta)}
  \widehat W(0,\eta',k) d\eta'
\end{equation}
and
\begin{equation}
  \label{eq:14b}
\begin{split}
 \lim_{\eps\to0+}\widehat{
    W}_\varepsilon^2(t,\eta,k) =\frac{|\nu(k)|^2}{4|\bar\om'(k)|} \sum_{\iota=\pm}\int_{\bbR}\frac{d\eta}{  \la+i \om'(k)\eta }
\int_{\bbR}
\frac{\widehat{W}(0,\eta',\iota k) d\eta'}{
  \la+i \om'(\iota k)\eta'}.
\end{split}
\end{equation}

Putting together the limits of the solutions   for
$T\ge0$ and $\widehat W_0=0$, given by \eqref{eq:8}, and the solution
of the deterministic equation for $T=0$ and a non-vanishing $\widehat W_0$
 obtained by taking the sum of the limits of $\widehat{
    W}_\varepsilon^j(t,\eta,k)$, $j=0,1,2$  we conclude the formula
  for the limit as $\'eps\to0$, of Wigner function $\widehat{
    W}_\varepsilon(t,\eta,k)$, see \eqref{eq:20}, for
  $\hat\phi(t,k)$, given by \eqref{eq:10}, equals:
\begin{equation}
  \label{eq:22}
  \begin{split}
    &W(t,y,k) 
    = 1_{[[0, \bar\omega'(k)t]]^c}(y) W (0,y - \bar \omega'(k) t ,k)\\
  & +p_+(k) 1_{[[0, \bar\omega'(k)t]]}(y)
    W (0,y - \bar \omega'(k) t ,k)\\
    & + p_-(k)
    1_{[[0, \bar\omega'(k)t]]}(y) W(0,-y + \bar\omega'(k) t,-k)
    + T  \fgeeszett(k)  1_{[[0, \bar\omega'(k)t]]}(y),
  \end{split}
\end{equation}
where the coefficient $\fgeeszett(k)$ is given by \eqref{eq:9} and 
\begin{equation}
  \label{eq:23}
\begin{split}
&   p_+(k):= \left|1 - \frac{\gamma \nu(k)}{2|\bar\omega'(k)|}\right|^2,
\quad
  p_-(k):= \left(\frac { \gamma|\nu(k)| }{2|\bar\omega'(k)|}\right)^2.
\end{split}
\end{equation}

By a direct inspection we can verify that $W(t,y,k)$, given by \eqref{eq:22}, solves
the transport equation
\begin{equation}
  \label{eq:6bis}
  \partial_t W(t,y,k) + \bar \om'(k) \partial_y W(t,y,k) = 0, \qquad y\neq 0,
\end{equation}
with the transmission/reflection and phonon creation boundary condition at $y=0$:
 \begin{equation}\label{feb1408}
\begin{split}
W(t,0^+,k)&=p_-(k)W(t,0^+, -k)+p_+(k)W(t,0^-,k)+{\frak g}(k)T,
\quad\hbox{ for $k\in\bbT_+$}\\
&\\
W(t,0^-, k)&=p_-(k)W(t,0^-,-k) + p_+(k)W(t,0^+, k) +{\frak
  g}(k)T,\quad \hbox{ for $k\in\bbT_-$.}
\end{split}
\end{equation}
{In Section
\ref{sec:prop-scatt-coeff} below we show that 
\begin{equation}
\label{012408-20}
p_+(k) +  p_-(k) + \fgeeszett(k) = 1.
\end{equation}
Coefficients $p_+(k)$ and $p_-(k)$ 
can be interpreted therefore as the probabilities of phonon
transmission and
reflection, respectively.
Since  $p_+(k)+p_-(k)=1-\fgeeszett(k)$, the coefficient
$\fgeeszett(k)$ is the phonon absorption probability
at the interface.}

\section{Harmonic chain with bulk conservative
  noise in contact with Langevin thermostat}
\label{sec:harmonic-chain-with}

\subsection{The model and the statement of the result} In \cite{ko19} we consider a stochastically perturbed chain of harmonic oscillators
thermostatted at a fixed temperature $T\ge 0$ at $x=0$. Its dynamics is
described by the system of It\^o stochastic differential equations
\begin{equation}
\label{eq:bas1}
\begin{split}
&d{\frak q}_{x}(t)={\frak p}_x(t)dt,\quad \quad x\in\bbZ,
\\
 &d{\frak p}_x(t)=\left[-(\alpha\star{\frak
   q}(t))_x-\frac{\eps\ga_0}{2}(\theta\star{\frak
   p}(t))_x\right]dt\\
&
+\sqrt{\eps\ga_0}\sum_{k=-1,0,1}(Y_{x+k}{\frak
   p}_x(t))dw_{x+k}(t) 
+\left(-\ga{\frak p}_0(t)dt+\sqrt{2\ga
    T}dw(t)\right)\delta_{0,x}.
\end{split}
\end{equation}
Here the coupling constants $(\al_x)_{x\in\bbZ}$ are as in
\eqref{011210}, 
\begin{equation}
\label{011210a}
Y_x:=({\frak p}_x-{\frak p}_{x+1})\partial_{{\frak p}_{x-1}}+({\frak p}_{x+1}-{\frak p}_{x-1})\partial_{{\frak p}_{x}}+({\frak p}_{x-1}-{\frak p}_{x})\partial_{{\frak p}_{x+1}}
\end{equation}
and
  $\left(w_x(t)\right)_{t\ge0}$, $x\in\bbZ$ with
  $\left(w(t)\right)_{t\ge0}$,  are i.i.d. one dimensional independent Brownian motions.
  In addition,
 $$
\theta_x=\Delta\theta^{(0)}_x:=\theta^{(0)}_{x+1}+\theta^{(0)}_{x-1}-2\theta^{(0)}_x
$$
with 
$$
 \theta^{(0)}_x=\left\{
 \begin{array}{rl}
 -4,&x=0\\
 -1,&x=\pm 1\\
 0, &\mbox{ if otherwise.}
 \end{array}
 \right.
 $$
Parameters  $\eps\gamma_0>0$, $\ga$ describe the strength of the
inter-particle and thermostat noises, respectively.
In what follows we shall assume that $\eps>0$ is small, that
corresponds to the low density hypothesis  that results in  atoms suffering finitely many
''collisions'' in a macroscopic unit of time  (the Boltzmann-Grad limit).
Although the noise  considered here 
is continuous we believe that the results extend to other type of conservative
noises, such as e.g. Poisson exchanges of velocities between nearest neighbor particles.

Since the vector field $Y_x$ is orthogonal both to a sphere ${\frak
  p}_{x-1}^2+{\frak p}_x^2+{\frak p}_{x+1}^2\equiv {\rm const}$ and 
plane ${\frak p}_{x-1}+{\frak p}_x+{\frak p}_{x+1}\equiv  {\rm const}$,
the inter-particle noise conserves locally the kinetic energy and
momentum.
{Because these conservation laws are common also for chaotic
hamiltonian system, this model has been used to understand
energy transport in presence of momentum conservation, see
\cite{bbjko16} and references there.}

The case without the Langevin thermostat, i.e. with $\gamma = 0$, was studied
in \cite{bos09}, where it is proved that 
\begin{equation}
  \label{eq:19}
{ W}_\varepsilon(t,y,k) 
  \mathop{\longrightarrow}_{\varepsilon\to 0} { W}(t,y,k) 
\end{equation}
where $W(t,y,k)$ is the solution of the transport equation
\begin{equation}
  \label{eq:bos6}
  \partial_t W(t,y,k) + \bar \om'(k) \partial_y W(t,y,k) =
  \gamma_0\int_{\bbT} R(k,k')\left(W(t,y,k') - W(t,y,k)\right) dk',
\end{equation}
where
\begin{equation}
  \label{eq:R}
  R(k,k') =32\sin^2(\pi k)\sin^2(\pi k')\left\{\sin^2(\pi k)\cos^2(\pi k')
      +\sin^2(\pi k') \cos^2(\pi k)\right\}.
  \end{equation}
{We have therefore 
\begin{equation}
\label{RR}
R(k)=\int_{\bbT} R(k,k')dk'=4\sin^2(\pi k)\big(1+3\cos^2(\pi k)\big).
\end{equation}}
  In \cite{ko19}  
  we have proved the following result.
\begin{theorem}
\label{thm010709-20}
 Suppose that $\ga_0,\gamma > 0$ and the initial data satisfies
 \eqref{011812aa}, 
 Then, 
 \begin{equation}
\label{040709-20}
\begin{split}
\lim_{\eps\to0+}  \int_0^{+\infty} dt \iint_{\bbT_{2/\varepsilon}\times\bbT}
{ W}_\varepsilon(t,y,k) 
G(t,y,k) dydk \\
= \int_0^{+\infty} dt \iint_{\bbR\times\bbT} { W}(t,y,k)G(t,y,k) dydk
\end{split}
\end{equation}  
for any $G\in C_0^\infty([0,+\infty)\times\bbR\times\bbT)$, where  the limiting Wigner $W(t,y,k)$  function satisfies \eqref{eq:bos6} for
  $(t, y,k)\in \bbR_+\times \mathbb R_*\times \mathbb T_*$, with the boundary conditions \eqref{feb1408},
  at the interface $y=0$.
\end{theorem}  

\subsection{A sketch of the proof of Theorem \ref{thm010709-20}}
\label{sec:scketch-proof}

Consider the wave function $\psi(t)$  that corresponds to the
dynamics \eqref{eq:bas1} via \eqref{011307}
and $\hat\psi(t)$
its Fourier transform.
In contrast with the situation described in Section
\ref{sec:evol-with-lang}  
(the case 
$\gamma_0 = 0$)  we  no longer have   an explicit expression for the solution of the
equation for $\hat\psi(t)$, see  \eqref{eq:sol1},
so we cannot proceed by a direct calculation of the
Wigner distributions as  in Sections \ref{sec:micdyn} and \ref{app:2}. 

{In order to close the dynamics of    the Fourier-Wigner
function, we shall  need all the components of 
the full covariance tensor 
of  the Fourier transform of the wave field.}
Define therefore the Wigner distribution tensor ${\bf W}_\eps(t)$,
as  a $2\times 2$-matrix tensor, whose
entries are distributions,  given by their respective  Fourier transforms
\begin{align}
\label{hbw}
&
\widehat \bW_\eps(t,\eta,k):=\left[\begin{array}{ll}
\widehat W_{\eps,+}(t,\eta,k)&\widehat Y_{\eps,+}(t,\eta,k)\\
\widehat Y_{\eps,-}(t,\eta,k)&\widehat W_{\eps,-}(t,\eta,k)
\end{array}\right], \quad (\eta,k)\in\bbT_{2/\eps}\times \bbT,
\end{align}
with
\begin{align*}
&
                 \widehat W_{\eps,+}(t,\eta,k):=\widehat W_{\eps}(t,\eta,k)=
                 \frac{\eps}{2}\bbE_\eps\left[\hat\psi\left(t/\eps,k+\frac{\eps\eta}{2}\right)
                 \hat\psi^\star\left(t/\eps,k-\frac{\eps\eta}{2}\right)\right],
\\
&
     \widehat Y_{\eps,+}(t,\eta,k):=\frac{\eps}{2}\bbE_\eps
     \left[\hat\psi\left(t/\eps,k+\frac{\eps\eta}{2}\right)
     \hat\psi\left(t/\eps,-k+\frac{\eps\eta}{2}\right)\right],
\\
&
\widehat Y_{\eps,-}(t,\eta,k):=\widehat
  Y_{\eps,+}^\star(t,-\eta,k),\quad 
\widehat W_{\eps,-}(t,\eta,k):=\widehat W_{\eps,+}(t,\eta,-k).
\end{align*}
By a direct calculation we show that 
the following energy bound is satisfied, see Proposition 2.1 of \cite{ko19}
\begin{equation}
\label{052709-18}
\sup_{\eps\in(0,1]}\frac{\eps}{2}\bbE_\eps\|\hat\psi(t/\eps)\|^2_{L^2(\bbT)}\le
\sup_{\eps\in(0,1]}\frac{\eps}{2}\bbE_\eps\|\hat\psi(0)\|^2_{L^2(\bbT)}
+\ga T t,\quad t\ge0.
\end{equation}
The above estimate implies in particular that
\begin{equation}
\label{030406-19}
\sup_{\eps\in(0,1]}\|  W_{\eps,+}\|_{L^\infty([0,\tau];{\cal
    A}')}<+\infty,\quad \mbox{for any }\tau>0,
\end{equation}
where
${\cal A}'$ is the dual to ${\cal A}$ 
- the Banach space obtained by the completion of ${\cal S}(\bbR\times\bbT)$  in the norm
\begin{equation}
\label{060805-19}
\|G\|_{\cal A}:=\int_{\bbR}\sup_{k\in\bbT}|\widehat G(\eta,k)|d\eta,\quad G\in {\cal S}(\bbR\times\bbT).
\end{equation}
Similar estimates hold also for the remaining entries of ${\bf
  W}_\eps(t)$. 

In consequence
$\left({\bf W}_{\eps}(\cdot)\right)$ is sequentially
$\star$-weakly compact in  
$L^\infty_{\rm loc}([0,+\infty),{\cal A}')$ and the problem of proving
its $\star$-weak convergence  reduces to the limit identification.

\subsubsection{The case of zero temperature at the thermostat}

 Suppose first that  $T=0$.
We can treat then the microscopic dynamics \eqref{eq:bas1} as a small
(stochastic) perturbation of the purely deterministic dynamics when all the terms
containing the noises $(w_x(t))$  and $(w(t))$ are omitted. Denote by
$\hat\phi^{(\eps)}\left(t,k\right)$ the Fourier transform of the wave
function corresponding to the latter dynamics, cf \eqref{011307a}. 
We  consider then the  respective Wigner distribution tensor
$
{\bf W}^{\rm un}_{\eps}(t,y,k)$ whose Fourier transform is given by an
analogue of \eqref{hbw}, where the wave function of the ''true''
(perturbed) dynamics $\hat\psi\left(t,k\right)$ is replaced
by $\hat\phi\left(t,k\right)$, which corresponds to the deterministic dynamics:
\begin{equation}
\label{eq:bas1det}
\begin{split}
d{\frak q}_{x}(t)&={\frak p}_x(t)dt,\quad \quad x\in\bbZ,
\\
d{\frak p}_x(t)&=\left[-(\alpha\star{\frak q}(t))_x
  -\frac{\eps\ga_0}{2}(\theta\star{\frak p}(t))_x\right]dt
-\ga{\frak p}_0(t) \delta_{0,x} dt.
\end{split}
\end{equation}

Denote by ${\cal  L}_{2,\eps}$
the Hilbert space made of the $2\times 2$ matrix valued 
distributions  on $\bbR\times\bbT$, such that the Fourier
transforms of their entries   belong to $L^2(\bbT_{2/\eps}\times\bbT) $.
The Hilbert norm on ${\cal L}_{2,\eps}$ is defined in an obvious
way using the $L^2$ norms of the Fourier transforms. 

Using the equations for the microscopic dynamics  of $\hat\phi\left(t,k\right)$
we conclude that
the tensor  ${\bW}^{\rm un}_{\eps}(t)$ can be described by
an ${\cal L}_{2,\eps}$ strongly continuous semigroup
$\Big({\frak W}_\eps^{\rm un}(t)\Big)$, i.e.
\begin{equation}
\label{frakWun}
{\bf W}^{\rm un}_{\eps}(t)={\frak W}_\eps^{\rm un}(t) \Big({\bf W}^{\rm un}_{\eps}(0)\Big),\quad t\ge0.
\end{equation}
Using a very similar argument to that used in the case of $\ga_0=0$
(remember no noise is present in the unperturbed dynamics) we can
prove, see Theorem 5.7 of  \cite{ko19}, that
\begin{theorem}
\label{main:thm-un}
Under the assumptions on the initial data made in\eqref{011812aa}
and \eqref{010709-20}, 
we have 
$$
\lim_{\eps\to0+}\langle G,{\bf W}^{\rm un}_{\eps}(t)\rangle =\langle
G,{\bf W}^{\rm un}(t)\rangle, \quad t\ge0,\,G\in {\cal S}(\bbR\times\bbT)
$$
where 
\begin{align*}
&
{\bf W}^{\rm un}(t,y,k)
=\left[\begin{array}{cc}
W^{\rm un}_{+}(t,y,k)&0\\
0&W^{\rm un}_{+}(t,y,-k)
\end{array}\right],\quad (y,k)\in \bbR\times\bbT,
\end{align*}
and 
\begin{equation}
  \label{eq:8p}
 \partial_tW^{\rm un}(t,y,k) + \bar\om'(k) \partial_y W^{\rm un}(t,y,k) = -\ga_0R(k)  W^{\rm un}(t,y,k), 
  \quad (t,y,k)\in\bbR_+\times \bbR_*\times\bbT_*,
\end{equation}
 with the interface conditions
\eqref{feb1408} for $T=0$. 
\end{theorem}
Similarly to \eqref{eq:6bis} equation \eqref{eq:8p}
 can be  solved explicitly and we obtain $W^{\rm un}(t)
=  {\frak W}^{\rm un}_t(W ^{\rm un} (0))
$, where 
$
W^{\rm un}(t,y,k)
= 
e^{-\ga_0R(k)t}{\tilde W}^{\rm un}(t,y,k)
$
and ${\tilde W}^{\rm un}(t,y,k)$ is given by \eqref{eq:22}.
Consider a semigroup defined by
\begin{equation}
\label{010304s}
{\frak W}^{\rm un}_t(W ^{\rm un} (0))\left(y,k\right)
:=  
W^{\rm un}(t,y,k).
\end{equation}
One can show that $\left({\frak
    W}^{\rm un}_t\right)_{t\ge0}$ forms a strongly continuous
semigroup of contractions on any $L^p(\bbR\times\bbT)$, $1\le
p<+\infty$.

We can use the semigroup ${\frak W}_\eps^{\rm un}(t)$ to write a
Duhamel type equation for   
$${\bf w}_\eps(\la)=
\left[\begin{array}{ll}
\widehat w_{\eps,+}(\la,\eta,k)&\widehat y_{\eps,+}(\la,\eta,k)\\
\widehat y_{\eps,-}(\la,\eta,k)&\widehat w_{\eps,-}(\la,\eta,k)
\end{array}\right]=\int_0^{+\infty}e^{-\la t}{\bf W}_\eps(t)dt
$$ 
-  the Laplace transform of the Wigner tensor of  \eqref{hbw} defined for ${\rm Re}\,\la>\la_0$ and some sufficiently large $\la_0>0$. It reads
\begin{equation}
\label{020805-19a}
{\bf w}_\eps(\la)  =\tilde{\frak W}_\eps^{{\rm
    un}}(\la) {\bf  W}_{\eps}(0)+\frac{\ga_0}{2} \tilde{\frak W}_\eps^{\rm
  un}\left(\la\right){\bf v}_\eps\left(\la\right),\quad {\rm Re}\, \la>\la_0.
\end{equation}
Here
\begin{equation}
\label{010701-19}
{\frak W}_\eps^{{\rm
    un}}(\la ):=\int_0^{+\infty}e^{-\la t}{\frak W}_\eps^{\rm un}(t)dt,
\end{equation}
and 
${\bf v}_\eps\left(\la\right):={\frak R}_\eps {\bf w}_\eps(\la)$.
The operator ${\frak R}_\eps$ acts on  $2\times2$ matrix valued 
${\bf w}$  whose entries belong to ${\cal L}_{2,\eps}$  and whose  Fourier transform (in $y$)  is given by
$$
\widehat{\bf w}(\eta,k):=\left[\begin{array}{ll}
\widehat w_{+}(\eta,k)&\widehat y_{+}(\eta,k)\\
\widehat y_{-}(\eta,k)&\widehat w_{\eps,-}(\eta,k)
\end{array}\right], \quad (\eta,k)\in\bbT_{2/\eps}\times \bbT
$$ 
as follows.
 The Fourier transform  $\widehat{{\frak R}_\eps{\bf w}}
$
 have entries of the form
\begin{align*}
\pm \int_{\bbT}r_0(k,k',\eps \eta)
\left[\widehat w_{\eps,+}(\eta,k')+\widehat w_{\eps,-}(\eta,k')-\widehat y_{\eps,+}(\eta,k')-\widehat y_{\eps,-}(\eta,k')\right]dk'.\nonumber
\end{align*}
Here $r_0(k,k',\eps \eta)$ is a scattering kernel that satisfies
$R(k,k')=r_0(k,k',0) +r_0(k,-k',0)$, with $R(k,k')$ given by
\eqref{eq:R}. 
Suppose now that we test  both sides of \eqref{020805-19a} against a
$2\times 2$-matrix valued  smooth   function ${\bf G}$ whose entries have compactly supported  Fourier transforms in the $y$ variable, say in the interval $[-K,K]$ for some $K>0$.  
Denote by $\Big( {\frak W}_{\eps}^{{\rm
    un}}(\la)\Big)^*$ and ${\cal R}_{\eps}^*$ the adjoints of the respective operators in ${\cal L}_{2,\eps}$.
    
We already know that from any sequence $ \Big({\bf w}_{\eps_n}  (\la)\Big)$, where $\eps_n\to0+$ we can choose a subsequence, that will be denoted by the same symbol, converging $\star$-weakly in ${\cal A}'$ to some ${\bf w} (\la)$.
Using Theorem \ref{main:thm-un} and the strong convergence of the sequence $1_{[-K,K]}(\eta)
{\cal R}_{\eps_n}^* \big(  {\frak W}_{\eps_n}^{{\rm
    un}}(\la )\big)^* {\bf G}$, $n\to+\infty$ in $L^2(\bbR\times\bbT)$  we can prove
the following, see Theorem 5.7 of \cite{ko19}. 
\begin{theorem}
\label{main:thm2}
Suppose that 
 ${\bf W}$ is the $\vphantom{1}^\star$-weak limit of  $\left({\bf
    W}_{\eps_n}\right)$ in $\left(L^1([0,+\infty);{\cal A})\right)'$ for some sequence
$\eps_n\to0+$. Then, it has to be of the form
\begin{equation}
\label{010702-19}
{\bf W}(t,y,k)
=\left[\begin{array}{cc}
{W}(t,y,k)&0\\
0&{W}(t,y,-k)
\end{array}\right],\quad (y,k)\in L^2(\bbR\times\bbT),
\end{equation}
where ${W}(t,y,k)$ satisfies the equation
\begin{equation}
\label{integral-bis}
W(t)={\frak W}^{\rm un}_t\Big(W(0)\Big)+\ga_0\int_0^t{\frak W}^{\rm
  un}_{t-s}\Big({\cal R} {W}_s\Big)ds,
\end{equation}
and ${\cal R}:L^2(\bbR\times\bbT)\to L^2(\bbR\times\bbT)$ is given by 
\begin{equation}
\label{cRW}
{\cal R}F(y,k):=  \int_{\bbT}R(k,k')
F\left(y,k'\right) dk',\quad (y,k)\in\bbR\times \bbT,\quad F\in L^2(\bbR\times \bbT).
\end{equation} 
\end{theorem}
The convergence claimed in Theorem \ref{thm010709-20} is then a direct consequence of
Theorems \ref{main:thm-un} and \ref{main:thm2}.

It turns out that the microscopic
evolution of the Wigner transform  given by \eqref{hbw}
allows us to define a strongly continuous semigroup on
$L^2(\bbT_{2/\eps}\times\bbT)$ by letting ${\frak
  W}_\eps(t)\Big({\bf W}_\eps(0)\Big):= {\bf W}_\eps(0)$. The norms of
the semigroups $L^2(\bbT_{2/\eps}\times\bbT)$ stay bounded with
$\eps\in(0,1]$, see Corollary 4.2 of \cite{ko19}.
Using Theorems \ref{main:thm-un} and \ref{main:thm2} we can show
therefore that for any ${\bf
  G}\in L^1([0,+\infty),{\cal A})$ we have
\begin{equation}
\label{121206-19}
\lim_{\eps\to0+}\int_0^{+\infty}\langle {\frak
  W}_\eps(t){\bf W}_\eps(0),{\bf G}(t)\rangle
dt=\int_0^{+\infty}\langle {\frak
  W}_t{\bf W}(0),{\bf G}(t)\rangle dt,
\end{equation}
where ${\frak
  W}_t{\bf W}(0):={\bf W}(t)$ is given by \eqref{010702-19}.

\subsubsection{The case of positive temperature at the thermostat}

Finally we consider the case $T>0$. Suppose that $\chi\in C^\infty_c(\bbR)$ is an arbitrary real valued,
even function satisfying   
\begin{equation}
\label{011406-19}
\chi(y)=\left\{\begin{array}{ll}
1,&\mbox{ for }|y|\le 1/2, \\
0,&\mbox{ for }|y|\ge 1,\\
\mbox{belongs to }[0,1],& \mbox{ if otherwise.}
\end{array}\right.
\end{equation}
Then  its Fourier transform $\widehat \chi\in {\cal S}(\bbR)$. Let 
 $\widehat \chi_\eps\in C^\infty(\bbT_{2/\eps})$ be given by 
$$
\widehat \chi_\eps(\eta):=\sum_{n\in\mathbb Z}\widehat
\chi\left(\eta+\frac{2n}{\eps}\right),\quad \eta\in\bbT_{2/\eps}.
$$
and
$$
\widehat {\bf V}_\eps(t,\eta,k):=\widehat {\bf W}_\eps(t,\eta,k)-T \widehat
\chi_\eps(\eta){\bf I}_{2},
$$
where ${\bf I}_2$ is the $2\times 2$ identity matrix.
In fact, ${\bf V}_\eps(t)$ is a solution of the equation
\begin{equation}
\label{111206-19}
{\bf V}_\eps(t)={\frak W}_\eps(t){\bf V}_\eps(0)+\int_0^t {\frak
  W}_\eps(s)\Big({\bf F}_\eps\Big) ds,
\end{equation}
where
$$
\widehat {\bf F}_\eps(\eta,k):=- \frac{iT\widehat
\chi_\eps(\eta)}{\eps}\left[\om\left(k+\frac{\eps \eta}{2}\right)-\om\left(k-\frac{\eps \eta}{2}\right)\right]  \left[\begin{array}{cc}
 1&0\\
 0&-1
 \end{array}\right].
$$
Using the convergence of \eqref{121206-19} we conclude that 
\begin{equation}
\label{121206-19a}
\lim_{\eps\to0+}\int_0^{+\infty}\langle {\bf V}_\eps(t),{\bf G}(t)\rangle dt=\int_0^{+\infty}\langle {\bf V}(t),{\bf G}(t)\rangle dt,
\end{equation}
where
$$
{\bf V}(t,y,k)= 
\left[\begin{array}{cc}
 V(t,y,k)&0\\
 0&V(t,y,-k)
 \end{array}\right]
$$
and
\begin{equation}
\label{111206-19a}
V(t,y,k):=W(t,y,k)+T\int_0^t {\frak
  W}_s\Big(\bar\om'(k)\chi'(y)\Big)ds.
\end{equation}
We can identify therefore $W(t,y,k)$ with the solution of the equation
\begin{equation}
\label{021406-19}
W(t,y,k):={\frak W}_t(\widetilde W_0)(y,k)+\int_0^t {\frak
  W}_s(F)(y,k)ds+T\chi(y),\quad (t,y,k)\in \bar\bbR_+\times\bbR\times\bbT.
\end{equation}
Here
\begin{equation}
\label{eqF}
F(y,k):=-T\bar\om'(k)\chi'(y),\quad \widetilde W_0(y,k):= W_0(y,k)-T\chi(y).
\end{equation}
 This ends the proof of Theorem \ref{thm010709-20} for an arbitrary $T\ge0$.

\section{Diffusive and Superdiffusive
  limit from the kinetic equation
    with boundary thermostat}
\label{sec:drodyn-limit-from}

We are now interested in the space-time rescaling of the solution of the equation
\eqref{eq:bos6} with the boundary condition \eqref{feb1408}.
We should distinguish here two cases:
\begin{itemize}
\item the \emph{optical chain}, when $\omega(k) \sim k^2$ for small $k$,
  \item the \emph{acoustic chain}, when $\omega(k) \sim |k|$ for small $k$.
\end{itemize}
In the optical chain, the long-wave phonons (corresponding to small $k$) have a small velocity,
consequently even if the bulk scattering rate is small ($R(k) \sim k^2$),
they still have time to diffuse. {In fact, all other phonons
(i.e. those corresponding to other $k$) have their non-trivial
contribution to the diffusive limit.}

In the acoustic chain the long-wave phonons move with the speed that is
bounded away from  $0$ and  rarely  scatter. Therefore, they are responsible
for a superdiffusion of the Levy type arising in the macroscopic
limit. {In the superdiffusive time-scale all other phonons
(corresponding to non-vanishng $k$) do not yet move, their
contribution to the asymptotic limit is therefore negligible.}


\subsection{The optical chain: diffusive behavior. }
\label{sec:optic-chain:-diff}

This case was studied in \cite{bko19}. The diffusive rescaling of the solution of
\eqref{eq:bos6} is defined by
\begin{equation}
  \label{eq:21}
  W^\delta(t,y,k) = W(t/\delta^2, y/\delta, k),
\end{equation}
with an initial condition that varying in the \emph{macroscopic} space scale
\begin{equation}
  \label{eq:25}
  W^\delta(0,y,k) = W_0(y,k). 
\end{equation}
We assume here that $W_0(y,k) = T + \tilde W_0(y,k)$, with
$\tilde W_0\in L^2(\mathbb R\times \mathbb T)$.
This rescaled solution solves
\begin{equation}
  \label{eq:bos6d}
  \partial_t W^\delta(t,y,k) + \frac 1\delta \bar \om'(k) \partial_y W^\delta(t,y,k) =
  \frac{\gamma_0}{ \delta^{2}}
  \int_{\bbT} R(k,k')\left(W^\delta(t,y,k') - W^\delta(t,y,k)\right) dk',
 \end{equation}
 with the boundary condition \eqref{feb1408} in $y=0$.

 In \cite{bko19} it is proven that, for any test function
 $\varphi(t,y,k) \in C^{\infty}_0( [0,+\infty)\times\mathbb R\times \mathbb T)$,
 \begin{equation}
   \label{eq:27}
   \lim_{\delta\to 0} \int_0^{+\infty} dt \iint_{\mathbb R\times \mathbb T}  W^\delta(t,y,k)
   \varphi(t,y,k)\; dy\; dk \ =\
   \int_0^{+\infty} dt \iint_{\mathbb R\times \mathbb T}  \rho(t,y) \varphi(t,y,k)\; dy\; dk,
 \end{equation}
 where $\rho(t,y)$ is the solution of the heat equation
 \begin{equation}
   \label{eq:26}
   \begin{split}
     \partial_t \rho(t,y) &= D \partial_y^2 \rho(t,y), \qquad y\neq 0,\\
     \rho(t,0) &= T,\qquad  \forall t>0, \\
     \rho(0,y) &= \rho_0(y) := \int_{\mathbb T}  W_0(y,k) dk.
 \end{split}
\end{equation}
The diffusion coefficient is given by
\begin{equation}
  \label{eq:D}
  D : = \frac {1}{\gamma_0} \int_{\mathbb T}  \frac{\bar\omega'(k)^2}{R(k)} dk.
\end{equation}
Notice that, under the condition of the optical dispersion relation,
$D<+\infty$.
The proof in \cite{bko19} follows a classical Hilbert expansion
method, with a modification needed to account for  the boundary condition.

Intuitively, the result can be explained in the following way:
phonons of all frequencies  behave diffusively, under the scaling
they converge to Brownian motions with diffusion $D$, that has continuous path.
As they get close to the thermostat boundary, they cross it many times till they
get absorbed with probability 1 in the macroscopic time scale.
Consequently there is no (macroscopic)
trasmission of energy from one side to the other.
Phonons are created with intensity T,
and this explain the value at the boundary $y=0$.

\subsection{The acoustic chain: superdiffusive behavior}
\label{sec:acoust-chain:-superd}

This limit was studied in \cite{kor19}, while the case without thermostat
had been previously considered in \cite{jko09}.
In a one dimensional acoustic chain, long wave phonons (small $k$)
move with finite velocities but still scatter very rarely.
Consequently these longwaves phonons on the microscopic scale
move ballistically with some rare scattering of their velocities. Under the
superdiffusive rescaling $\delta^{-3/2} t, \delta^{-1} y$ they converge to
corresponding Levy processes, generated by the fractional laplacian
$-|\Delta|^{3/4}$. The effect of the thermal boundary is more complex
than in the diffusive case, as now the phonons have a positive probability
to cross the boundary without absorption and jump at a macroscopic distance
on the other side. This causes a particular boundary condition for the
fractional laplacian at the interface $y=0$,
that we explain below.
Let us define the fractional laplacian $-|\Delta|^{3/4}$, admitting an interface  value
$T$, with absorption ${\frak g}_0$, transmission $p_+$ and reflection $p_-$,
  as  the $L^2$  closure of the 
singular integral operator   
\begin{equation}\label{apr204}
\begin{aligned}
  \Lambda_{3/4} F(y) = &{\rm p.v.}\,\int_{yy'>0}q(y-y')[F(y')-F(y)]dy' \\
  & +  {\frak g}_0[T-F(y)] \int_{yy'<0}q(y-y')dy'\\
  &+ p_-\int_{yy'<0}q(y-y')[F(-y')-F(y)]dy'\\
  &+ p_+\int_{yy'<0}q(y-y')[F(y')-F(y)]dy',\quad y\not=0,\,F\in C_0^\infty(\bbR),
\end{aligned}
\end{equation}
{where, cf \cite[Theorem 1.1 e)]{kw},
\begin{equation}
  \label{eq:29}
  q(y) = \frac{c_{3/4}}{|y|^{5/2}}, \qquad
  c_{3/4} = \frac{2^{3/2} \Gamma(5/4)}{\sqrt{\pi}|\Gamma(-3/4)|}=\frac{3}{2^{5/2}\sqrt{\pi}}.
\end{equation}
The first integral appearing in the right hand side of \eqref{apr204} is
understood in the principal value (p.v.) sense. The choice of constant $c_{3/4} $ is made in such a way that the ''free'' fractional laplacian, defined by the kernel $q(\cdot)$, coincides with the 
definition using the ''usual'' Fourier symbol, see \cite[Theorem 1.1 a)]{kw}.
To define  $\Lambda_{3/4} F(0)$, note that, due to the
fact that  ${\frak
  g}_0>0$, the
finiteness of the second integral forces the condition $F(0)=T$ on any function belonging to the domain of the generator. We can define   $\Lambda_{3/4} F(0)$ using \eqref{apr204} for any continuous function that satisfies $F(0)=T$,  for which the integrals appearing in the right hand side (without the principal value) converge.}

Notice that in the case without thermal interface, ${\frak g}_0 = 0,
p_- = 0,  p_+ = 1$, and
we recover the usual ''free'' fractional laplacian on the real line.
The absorption, transmission and reflection coefficients that arise here are
given by
\begin{equation}
  \label{eq:28}
  {\frak g}_0 = \lim_{k\to 0} {\frak g}(k), \quad
  p_\pm = \lim_{k\to 0} p_\pm(k).
\end{equation}
For the nearest neighbor   acoustic chain, with the dispersion
relation $\om(k):=\om_a|\sin(\pi k)|$ (cf \eqref{om-p2})  it turns
out that, see \eqref{eq:23a},
  \begin{equation}
  \label{eq:23b}
\begin{split}
   p_+ =\left(\farc{\om_a}{\om_a+\gamma}\right)^2,
\qquad
  p_-(k):= \left(\farc{\ga }{\om_a+\gamma}\right)^2
\qquad \fgeeszett_0= \farc{2\ga \om_a}{(\om_a+\gamma)^2}.
\end{split}
\end{equation}
The rescaled solution of the kinetic equation, see \eqref{eq:bos6}, is defined now by
\begin{equation}
  \label{eq:21sd}
  W^\delta(t,y,k) = W(t/\delta^{3/2}, y/\delta, k)
\end{equation}
In \cite{kor19} it is proven that for any $t>0$ and $\varphi\in C_0^\infty(\bbR\times\bbT)$
 \begin{equation}
   \label{eq:27sd}
   \lim_{\delta\to 0} \iint_{\mathbb R\times \mathbb T}  W^\delta(t,y,k)
   \varphi(y,k)\; dy\; dk \ =\
 \iint_{\mathbb R\times \mathbb T}  \rho(t,y) \varphi(y,k)\; dy\;
 dk,\quad 
\end{equation}
where $\rho(t,y)$ is the solution of
\begin{equation}
  \label{eq:31}
  \partial_t \rho(t,y) = \hat c \Lambda_{3/4} \rho(t,y),
\end{equation}
where 
 {\begin{equation}
\label{hatc}
\hat c:={\frac{\pi^2\om_a^{3/2}}{ (2^5\ga_0)^{1/2}}}\,
 \int_{0}^{+\infty}\frac{(1 -\cos \la)d\la}{\la^{5/2}}=\left({\frac{\pi^5\om_a^{3}}{6\ga_0}}\right)^{1/2},
\end{equation}}
cf \cite[formula 3.762, 1, p. 437]{GR}

The proof of \eqref{eq:27sd}, presented in  \cite{kor19}, is based on the probabilistic
representation of the phonon trajectory process associated with  the
kinetic equation \eqref{eq:bos6}.
It is shown that  superdiffusively scaled trajectories of the process
converge in law to those of a Levy process,
with corresponding probabilities to be absorbed, transmitted or reflected
when crossing $y=0$, with a creation in the same point (its generator
is given by \eqref{apr204}).

{
  \begin{remarknn}
Notice that the convergence in \eqref{eq:27sd} holds for every time {$t> 0$},
while in the diffusive case it is only weakly in time (cf \eqref{eq:27}).
The explanation comes from different methods adopted  in the
respective proofs. {The proof of \eqref{eq:27sd} is of
  probabilistic nature, and
uses the fact that  the corresponding limiting
transmitted/reflected/absorbed  process jumps over
the thermostat  interface only finitely many times  
before being absorbed}. On the other hand, the proof of \eqref{eq:27} is analytic,
and it would be difficult to establish, by a probabilistic method, a result for every time, since the
corresponding Brownian motion crosses the thermostat infinitely
many times before being absorbed by it. 
\end{remarknn}
  }

\section{Perspectives and open problems}
\label{sec:pers-open-probl}

\subsection{Direct hydrodynamic limit}
\label{sec:direct-hydr-limit}

The results  presented in the previous sections are obtained
in the typical two-step procedure: we first take a kinetic limit (rarefied collisions)
and obtain a kinetic equation with a boundary condition for the thermostat,
next we rescale (diffusively or superdiffusively) this equation getting
a diffusive or superdiffusive equation with an appropriate boundary condition.

It would be interesting to obtain a direct hydrodynamic limit, rescaling diffusively
or superdiffusively the microscopic dynamics, without rarefaction
of the random collision in the bulk. This means considering the evolution
equations \eqref{eq:bas1} with $\varepsilon = 1$, then setting a scale parameter
$\delta$ (that does not appear in the evolution equations) and define
the Wigner distribution by
\begin{equation}
\label{wignerHL}
\langle G,W^{(\delta)}(t)\rangle:=\frac{\delta}{2}
\sum_{y,y'\in\bbZ}\int_{\bbT} e^{2\pi ik(y'-y)}\bbE \left[\psi_y \left(\frac{t}{\delta^\alpha}\right)
\left(\psi_{y'}\right)^*\left(\frac{t}{\delta^\alpha}\right)\right]
G^*\Big(\delta\frac{y+y'}{2},k\Big)dk. 
\end{equation}
with $\alpha = 2$,  or $\alpha= 3/2$ in the diffusive, or
superdiffusive case, respectively.
Then one would like to show that, in some sense,
\begin{equation}
  \label{eq:30}
  W^{(\delta)}(t,y,k) \mathop{\longrightarrow}_{\delta\to 0} \rho(t,y),
\end{equation}
where $\rho(t,y)$ is solution of \eqref{eq:26} or \eqref{eq:31},
depending on the scaling. In absence of a thermostat, this has been proved in \cite{jko15}.

\subsection{More thermostats}
\label{sec:more-thermostats}

In non-equilibrium statistical mechanics it is always interesting to put the system
in contact with a number of heat baths atvarious temperatures. If, in
the case of dynamics
defined by \eqref{eq:bas2} or by \eqref{eq:bas1},
we add another Langevin thermostat at the site $[\epsilon^{-1} y_0]$
with $y_0\not=0$, at a 
temperature $T_1$, we expect to obtain the same kinetic equations with added boundary conditions
at the point $y_0$ analogous to \eqref{feb1408} but of course the
phonon production rate ${\frak g}(k)T_1$.
The difficulty in constructing the proof, lies in the fact that we
no longer have an explicit formula for a solution in the case 
the inter-particle scattering is absent, that has been quite essential in our argument.


\subsection{Poisson thermostat}
\label{sec:poisson-thermostat}

A different model for a heat bath at temperature $T$ is given by a
renewal of the velocity $p_0(t)$ at random times given by a Poisson process
of intensity $\gamma$:  each time the Poisson clock rings, the velocity
is renewed with value chosen with a Gaussian distribution of variance $T$,
independently of  anything else. This mechanism represents
the interaction with an infinitely extended reservoir of independent particles
in equilibrium at temperature T and uniform density.

From a preliminary calculation (cf \cite{koP}) it seems that the scattering rates
in the high frequency-kinetic limit 
are different, implying their dependence  on the microscopic model of the
thermostat. Obviously,  in the hydrodynamic limit, diffusive or superdiffusive,
we expect that there boundary conditions will not
depend anymore on the microscopic model of the thermostat.

\begin{acknowledgement}
  TK was partially supported by the NCN grant 2016/23/B/ST1/00492, 
SO by the French Agence Nationale Recherche grant LSD ANR-15-CE40-0020-01.
\end{acknowledgement}


\section{Appendix A: Properties of the scattering coefficients}
\label{sec:prop-scatt-coeff}

\subsection{Some properties of the scattering coefficient for a
  unimodal dispersion relation}  
Recall that $\nu(k)$ is defined by \eqref{nu}. From  (\ref{eq:2}), we have
$$
 \lim_{\eps\to0}\tilde J(\eps -i\om(k))
=iG\big(\om(k)\big)+iH\big(\om(k)\big) ,\quad \mbox{for }\om(k)\not=0
$$
where
\begin{equation}
\label{G-H}
G(u):=\int_{\bbT_+}\frac{ d\ell}{u+\om(\ell)},
\qquad
H(u):= \farc 12\lim_{\eps\to0}\int_{\bbT}\frac{
  d\ell}{i\eps+u-\om(\ell)}
.\end{equation}
If $\om(k)=0$, then $k=0$ and, according to \eqref{eq:2}, 
\begin{equation}
\label{012608-20}
 \lim_{\eps\to0}\tilde J(\eps)
=\frac{\pi}{|\om'(0+)|}.
\end{equation}

If the dispersion relation $\om(k)$ is unimodal and $\omega_{\min}:=\om(0)$, $\omega_{\max}:=\om(1/2)$,
then we can write $H(\omega(k))=H^r(\omega(k))+
i H^i(\omega(k)),$
with $H^r(u)$, $H^i(u)$ real valued functions equal
\begin{equation}
\label{Hr}
H^r(u) :=\lim_{\eps\to0}
\int_{\omega_{\min}}^{\omega_{\max}}\farc{(u-v)dv}{|\omega'(\omega_+^{-1}(v))|[\eps^2+(u-v)^2]}
\end{equation}
and
\begin{equation}
\label{Hi}
H^i(u) :=-\lim_{\eps\to0}
\int_{\omega_{\min}}^{\omega_{\max}}\farc{\eps dv}{|\omega'(\omega_+^{-1}(v))|[\eps^2+(u-v)^2]}=-\farc{\pi }{|\omega'(\omega_+^{-1}(u))|}.
\end{equation}
Here $\om_+^{-1}:[\omega_{\min}, \omega_{\max}]\to[0,1/2]$ is the
inverse of the increasing branch of  $\om(\cdot)$. 
For $u\in (\omega_{\min}, \omega_{\max})$ we can write
\begin{equation}
\label{Href}
H^r(u)
=\farc{1}{\omega'(\omega_+^{-1}(u))}\log\frac{\omega_{\max}-u}{u-\omega_{\min}}+
\int_{\omega_{\min}}^{\omega_{\max}}\farc{\Big[\Big(\om_+^{-1} \Big)'(v)-\Big(\om_+^{-1} \Big)'(u)\Big]dv}{u-v}.
\end{equation}
According to \eqref{tg} and \eqref{nu} 
\begin{equation}\label{mar1524}
\nu(k)=
\left\{
\begin{array}{ll}
\{1-\ga H^i(\omega(k))
  +i\gamma{[G(\omega(k))+H^r(\omega(k))]}\}^{-1},&\mbox{ if
                                                   }\om(k)\not=0,\\
&\\
\dfrac{2|\bar\om'(0+)|}{2|\bar\om'(0+)|+\ga},&\mbox{ if
                                                   }\om(k)=0.
\end{array}
\right.
\end{equation}
Summarizing, from the above argument we conclude the following.
\begin{theorem}
\label{thm01}
For a unimodal dispersion relation $\om(\cdot)$ the following are true: 
\begin{itemize}
\item[i)] we have
\begin{equation}
\label{feb1402a}
|\nu(k)|\le  \frac{2|\bar\om'(k)|}{\ga+2|\bar\om'(k)|},\quad k\in\bbT,
\end{equation}
\item[ii)]  if $k_*$ is such that $\om'(k_*)=0$, then
\begin{equation}
\label{feb1402c}
\lim_{k\to k_*}\nu(k)=0\quad\mbox{and}\quad \lim_{k\to k_*}\fgeeszett(k)=0,
\end{equation}
\item[iii)]  
\begin{equation}
\label{feb1402b}
{\rm
  Re}\,\nu(k)>0,\quad\mbox{for all }k\in\bbT\setminus\{0,1/2\},
\end{equation}
\item[iv)]  
\begin{equation}
\label{feb1402d}
p_+(k)>0\quad\mbox{and }p_-(k)<1\mbox{ for all }k\,\mbox{such
  that }\om'(k)\not=0
\end{equation}
and
\begin{equation}
\label{feb1402e}
p_-(k)>0\,\mbox{ for all }k\in\bbT\setminus\{0,1/2\},
\end{equation}
\item[v)]
we have the formula
\begin{equation}
\label{feb1402}
{\rm
  Re}\,\nu(k)=\left(1+\frac{\ga}{2|\bar\om'(k)|}\right)|\nu(k)|^2,\quad
k\in\bbT.
\end{equation}
\end{itemize}
\end{theorem}
\proof
Substituting into (\ref{mar1524}) from \eqref{Hr} and \eqref{Hi}
immediately yields \eqref{feb1402}. Estimate \eqref{feb1402b} follows directly from \eqref{mar1524}, 
formulas \eqref{G-H}, \eqref{Hi} and \eqref{Href}.

Statement \eqref{feb1402a} is a consequence of \eqref{Hi} and \eqref{mar1524}.
Part ii) follows from part i), cf
\eqref{eq:9}.
Estimates \eqref{feb1402d} follow directly from \eqref{feb1402a}, while  \eqref{feb1402e}
is a straightforward consequence of part iii), cf
\eqref{eq:23}.
$\Box$

\medskip

From part v) of Theorem \ref{thm01} we immediately conclude the
following.
\begin{corollary}
Suppose that $\nu(k)\not=0$ is real valued. Then,
\begin{equation}
\label{feb1402a}
\nu(k) =\frac{|\om'(k)|}{|\om'(k)|+\ga\pi}.
\end{equation}
\end{corollary}

\subsection{Proof of (\ref{012408-20})}  To conclude \eqref{012408-20} we invoke \eqref{eq:23}. Then, thanks to
\eqref{feb1402}, we can write
\begin{align*}
p_+(k)+p_-(k)+\fgeeszett(k)=1+\frac{\gamma 
  }{|\bar\omega'(k)|}\left[|\nu(k)|^2\left(1+\frac { \gamma }{2|\bar\omega'(k)|} \right)  -{\rm Re}\, \nu(k)\right]=1
\end{align*}
and \eqref{012408-20} follows.

\subsection{An example: scattering coefficient $\nu(k)$ for a
 nearest neighbor interaction harmonic chain - computation of $\tilde J(\la)$ using  contour integration}

Assume that $\om(k)$ is the dispersion relation of  a
 nearest neighbor interaction harmonic chain. We let
 $\al_{0}:=(\om_0^2+\om_a^2)/2$ and  $\al_{\pm1}:=-\om_a^2/4$, and
 $\om_0\ge0$, $\om_a>0$.
Then, see Section \ref{sec:micdyn}, 
\begin{equation}
\label{al}
\hat\al(k)=\frac{\om_0^2+\om_a^2}{2}-\frac{\om_a^2}{4}\left(e^{2\pi i k}+e^{-2\pi i k}\right)=\frac{\om_0^2}{2}+\om_a^2\sin^2(\pi k)
\end{equation}
and, according to \eqref{mar2602},
\begin{equation}
\label{om-p2}
\om(k):=\sqrt{\frac{\om_0^2}{2}+\om_a^2\sin^2(\pi k)}.
\end{equation}
Using the definition of  $\tilde J(\la)$, see \eqref{eq:2}, and
\eqref{al} for any $\la\in\mathbb C$ such that ${\rm Re}\,\la>0$ we can write 
\begin{equation}
\label{circ}
\begin{split}
\tilde J(\la)=\int_{\bbT}\frac{\la d\ell}{\la^2+\hat\al(\ell)}=-\frac{4\la}{\om_a^2}\int_{-1/2}^{1/2}\frac{e^{2\pi
    i\ell}d\ell}{e^{4\pi i\ell}-2W(\la) e^{2\pi i\ell}+1},
\end{split}
\end{equation}
where
\begin{equation}
\label{W}
W(\la) =1+\left(\frac{\om_0}{\om_a}\right)^2+2 \left(\frac{\la}{\om_a}\right)^2.
\end{equation}
Note that $W(\la)\in\mathbb C\setminus [-1,1]$, if ${\rm Re}\,\la>0$.

The expression for $\tilde J(\la)$ can be rewritten using the contour
integral over the unit circle $C(1)$ on the complex plane oriented
counterclockwise and 
\begin{equation}
\label{circ1}
\begin{split}
\tilde J(\la)=-\frac{2\la}{i\pi\om_a^2}\int_{C(1)}\frac{d\zeta}{\zeta^2-2W(\la) \zeta+1}.
\end{split}
\end{equation}


When $w\in\mathbb C\setminus [-1,1]$ the equation
$$
z^2-2wz+1=0
$$
has two roots. They are given by $\Phi_+,\Phi_-$, holomorphic functions
on $\mathbb C\setminus [-1,1]$, that are the inverse branches of the
Joukowsky function ${\frak J}(z)=1/2(z+z^{-1})$, $z\in \mathbb C$
taking values in $\bar{\mathbb D}^c$ and $\mathbb D$,
respectively. Here $\mathbb D:=[z\in\mathbb C:\,|z|<1]$ is the unit
disc. We have
 \begin{equation}
\label{ai} 
\lim_{\eps\to0+}\frac{1}{2}\Big(\Phi_+(a-\eps i)-\Phi_-(a-\eps
  i)\Big)=-i\sqrt{1-a^2},\quad \mbox{for }a\in[-1,1].
\end{equation}
Using the Cauchy formula for contour integrals, from \eqref{circ1} we obtain
\begin{equation}
\label{circ1a}
\tilde
J(\la)
=\frac{2\la}{\om_a^2 \Big(\Phi_+(W(\la))-\Phi_-(W(\la))\Big)}.
\end{equation}
For the dispersion relation $\om(k)$ given by \eqref{om-p2} and $\eps>0$ we have, cf \eqref{W},
$$
W(\eps-i\om(k))=\cos(2\pi k)+2 \left(\frac{\eps}{\om_a}\right)^2-4i \frac{\om(k)\eps}{\om_a^2}.
$$
As a result we get, cf \eqref{ai} and \eqref{circ1a},
$$
\lim_{\eps\to0+}\tilde
J(\eps-i\om(k))=\frac{2}{\om_a^2\sin(2\pi
  |k|)}\sqrt{\frac{\om_0^2}{2}+\om_a^2\sin^2(\pi k)}
$$
and the following result holds.
\begin{theorem}
\label{thm02}
For the dispersion relation given by \eqref{om-p2} we have
\begin{equation}
\label{nu1}
\nu(k)=\om^2_a\sin(2\pi|k|)\left\{\om^2_a\sin(2\pi|k|)+2\ga\sqrt{\frac{\om_0^2}{2}+\om_a^2\sin^2(\pi
    k)}\right\}^{-1},\quad k\in\bbT.
\end{equation}
In particular, if $\om_0=0$ (the acoustic case) we have, cf
\eqref{eq:9} and \eqref{eq:23},
\begin{equation}
\label{nu1a}
\nu(k)=\frac{\om_a\cos(\pi k)}{\om_a\cos(\pi k)+\ga},\quad k\in\bbT
\end{equation}
and 
\begin{equation}
  \label{eq:23a}
\begin{split}
   &p_+(k):=\left(\farc{\om_a\cos(\pi k) }{\om_a\cos(\pi k)+\gamma}\right)^2,
\qquad
  p_-(k):= \left(\farc{\ga }{\om_a\cos(\pi k)+\gamma}\right)^2\\
&
~~~~~~~~~~~~~~~~~~~~~~~~~~~\fgeeszett(k)= \farc{2\ga \om_a\cos(\pi k)}{(\om_a\cos(\pi k)+\gamma)^2},\quad k\in\bbT.
\end{split}
\end{equation}
\end{theorem}

\section{Appendix B: proofs of (\ref{eq:13b})
  and (\ref{eq:14b})}
\addcontentsline{toc}{section}{Appendix 2}
\label{app:2}

\subsection{Proof of  (\ref{eq:13b})}

Using \eqref{eq:1} and \eqref{null}  we can write
\begin{equation}
  \label{eq:5-det-2a}
\widehat{
    W}_\varepsilon^1(t,\eta,k)=-\frac{\gamma}{2}
 \left\{ {\cal I}\left(\frac{t}{\eps},\eta,k\right) +{\cal I}^*\left(\frac{t}{\eps},-\eta,k\right)\right\},
\end{equation}
where
\begin{equation}
  \label{eq:5-det-2}
  \begin{split}
 &{\cal I}(t,\eta,k) := \frac{\eps}{2}\int_{\bbT} dk' \bbE_\eps\Big[ \hat\psi\left(0, k-\frac{\eps\eta}{2}\right)^*
 \hat\psi\left(0, k'\right) \Big]
  \int_0^{t}
  e^{i[\omega(k-\frac{\eps\eta}{2}) t-\omega(k')s]} 
  \phi\left( t-s, k+\frac{\eps\eta}{2}\right) ds.
\end{split}
\end{equation}
The Laplace transform of ${\cal
  I}\left(\frac{t}{\eps},\eta,k\right) $ equals
\begin{equation}
  \label{eq:5-det-x}
  \begin{split}
 &\tilde I_\eps(\la,\eta,k):=\frac{\eps^2}{2}\int_0^{+\infty}e^{-\eps\la t}{\cal
  I}\left(t,\eta,k\right)dt\\
&
=\frac{\eps^2}{2}\int_0^{+\infty} e^{i\omega(k+\frac{\eps\eta}{2}) \tau}g(d\tau)\int_{\tau}^{+\infty}  e^{i[\omega(k')-\omega(k+\frac{\eps\eta}{2}) ]s}  ds
  \int_s^{+\infty}e^{-\{\eps\la  +i[
    \omega(k')-\omega(k-\frac{\eps\eta}{2})] \}t} dt\\
&
\times
 \int_{\bbT} dk' \bbE_\eps\Big[ \hat\psi\left(0, k-\frac{\eps\eta}{2}\right)^*
 \hat\psi\left(0, k'\right) \Big].
\end{split}
\end{equation}
Performing the integration over the temporal variables we conclude that
\begin{equation}
  \label{eq:5-det-y}
  \begin{split}
 &\tilde I_\eps(\la,\eta,k)
=\int_{\bbT}\bbE_\eps\Big[ \hat\psi\left(0, k-\frac{\eps\eta}{2}\right)^*
 \hat\psi\left(0, k'\right) \Big]
\\
&
\times 
\frac{\tilde g\left(\la\eps-i\omega(k-\frac{\eps\eta}{2})\right)dk'}{2\{\la+i\eps^{-1}[\omega(k-\frac{\eps\eta}{2})-\omega(k')
    ]\}\{\la+i\eps^{-1}[
    \omega(k+\frac{\eps\eta}{2})-\omega(k-\frac{\eps\eta}{2})]\} }
.
\end{split}
\end{equation}
Using \eqref{nu} we conclude that for any test function $G\in {\cal
  S}(\bbR\times\bbT)$
{
\begin{equation}
  \label{eq:5-det-z}
  \begin{split}
 &\int_{\bbR\times\bbT}\tilde I_\eps(\la,\eta,k)\hat G^*(\eta,k)d\eta dk
\\
&
\approx \int_{\bbR\times\bbT^2}\frac{\bbE_\eps\Big[ \hat\psi\left(0, k\right)^*
 \hat\psi\left(0, k'\right) \Big]\hat G^*(\eta,k)\nu(k)d\eta dk dk'}{2\{\la+i\eps^{-1}[
    \omega(k')-\omega(k)]\}\{\la+i\eps^{-1}[
    \omega(k+\eps\eta)-\omega(k)]\} },
\end{split}
\end{equation}
as $\eps\ll1$.
Changing variables $k:=\ell-\eps\eta'/2$ and $k':=\ell+\eps\eta'/2$ we
obtain that
\begin{equation}
  \label{eq:5-det-z}
  \begin{split}
 &\lim_{\eps\to0+}\int_{\bbR\times\bbT}\tilde I_\eps(\la,\eta,k)\hat G^*(\eta,k)d\eta dk
\\
&
=\int_{\bbR^2\times\bbT}\frac{\widehat{ W}(0,\eta',\ell)\hat G^*(\eta,\ell)\nu(\ell)d\eta d\eta' d\ell}{(\la+i\omega'(\ell)\eta')
    (\la+i\omega'(\ell)\eta)}.
\end{split}
\end{equation}
The limit of $\hat{
    w}_\varepsilon^1(\la,\eta,k)$ - the Laplace transform of  $\widehat{
    W}_\varepsilon^1(t,\eta,k)$ -  is therefore given by 
\begin{equation}
  \label{eq:13}
\hat{
    w}^1(\la,\eta,k)=- \frac{\gamma \text{Re} \nu(k)}{\la+i\omega'(k)\eta} \int_{\bbR}\frac{\widehat{ W}(0,\eta',k) d\eta'}{\la+i\omega'(k)\eta'}
 \end{equation}}
therefore

\begin{equation}
  \label{eq:13bb}
\lim_{\eps\to0+}\widehat{
    W}_\varepsilon^1(t,\eta,k)=-  \gamma \text{Re} \nu(k) e^{-i\omega'(k)t} \int_{\bbR}
  \frac{1-e^{-i\omega'(k)(\eta'-\eta) t}}{i\omega'(k) (\eta'-\eta)}
  \widehat W(0,\eta',k) d\eta'
\end{equation}
and, performing the inverse Fourier transform, \eqref{eq:13b} follows.

\subsection{Proof of (\ref{eq:14b})}

Concerning the term $\widehat{
    W}_\varepsilon^2(t,\eta,k)$, from the third formula of
  \eqref{eq:5-det}, \eqref{null} and \eqref{eq:1} we obtain 
\begin{equation}
\label{eq:tk02}
  \begin{split} 
&\widehat{
    W}_\varepsilon^2(t,\eta,k)=\frac{\gamma^2}{2}
 \left\{ {\cal J}\left(\frac{t}{\eps},\eta,k\right) +{\cal R}\left(\frac{t}{\eps},\eta,k\right)\right\},
\end{split}
\end{equation}
where
\begin{equation}
\label{eq:tk01}
  \begin{split} 
& {\cal J} (t,\eta,k):= \frac{\eps}{4}  \int_{[0,t]^2}
dsds'\int_{\bbT^2}d\ell d\ell'
  \phi\left(t-s,k-\frac{\eps\eta}{2}\right)^*
  \phi\left(t-s',k+\frac{\eps\eta}{2}\right)\\
&
\times e^{i[\omega(\ell) s-\omega(\ell') s']} \bbE_\eps\left[
  \hat\psi(0,\ell) ^* \hat\psi(0,\ell')\right],\\
& {\cal R} (t,\eta,k):= \frac{\eps}{4} \int_{[0,t]^2}
dsds'\int_{\bbT^2}d\ell d\ell'
  \phi\left(t-s,k-\frac{\eps\eta}{2}\right)^*
  \phi\left(t-s',k+\frac{\eps\eta}{2}\right)\\
&
\times e^{-i[\omega(\ell) s-\omega(\ell') s']} \bbE_\eps\left[ \hat\psi(0,\ell) \hat\psi(0,\ell')^*\right].
\end{split}
\end{equation}
A simple computation shows that
\begin{equation}
\label{eq:J}
  \begin{split} 
& {\cal J} (t,\eta,k)= \frac{\eps}{4} \int_{[0,t]^2}dsds'\int_0^s  g(d\tau)g(d\tau')
\int_{\bbT^2}d\ell d\ell'  \bbE_\eps\left[
  \hat\psi(0,\ell) ^* \hat\psi(0,\ell')\right]
 \\
&
\times
e^{i[\om(k-\eps\eta/2)(s-\tau)-\omega(k+\eps\eta/2) (s'-\tau')]}
e^{-i[\omega(\ell')(t-s')-\omega(\ell)(t-s)]}.
\end{split}
\end{equation}
The respective Laplace transform equals
\begin{equation}
\label{eq:tJ}
  \begin{split} 
&\tilde{\cal J}_\eps (\la,\eta,k):=\eps\int_0^{+\infty}e^{-\eps\la \tau_0}
{\cal J} (\tau_0,\eta,k)d \tau_0 \\
&
=\eps\int_0^{+\infty}\int_0^{+\infty}\delta(\tau_0-\tau_0') e^{-\eps\la \tau_0/2}e^{-\eps\la \tau_0'/2}
{\cal J} (\tau_0,\eta,k){\cal J} (\tau_0',\eta,k)d \tau_0 d \tau_0'\\
&
= \frac{\eps^2}{4} \int_{\bar\bbR_+^4\times
  \bar\bbR_+^4 } d\tau_{0,2} g(d\tau_3) d\tau_{0,2}'g(d\tau_3')
\int_{\bbT^2}d\ell d\ell'  
 \\
&
\times 
\delta(\tau_0-\tau_0')\delta\left(\tau_0-\sum_{j=1}^3\tau_j
\right)\delta\left(\tau_0'-\sum_{j=1}^3\tau_j'\right) e^{-\eps\la \sum_{j=0}^3\tau_j/4}e^{-\eps\la \sum_{j=0}^3\tau_j'/4} \\
&
\times e^{i[\om(k-\eps\eta/2)\tau_2+\omega(\ell)\tau_1]}
e^{-i[\om(k+\eps\eta/2)\tau_2'+\omega(\ell')\tau_1']}\bbE_\eps\left[
  \hat\psi(0,\ell) ^* \hat\psi(0,\ell')\right].
\end{split}
\end{equation}
Here, for abbreviation sake we write $d\tau_{0,3}=d\tau_0d\tau_1d\tau_2$ and
likewise for the prime variables.  
Using the identity
$\delta(t)=(2\pi)^{-1}\int_{\bbR}e^{i\beta t}d\beta$ and integrating
out the $\tau$ and $\tau'$ variables we obtain
\begin{equation}
\label{eq:tJ1}
  \begin{split} 
&\tilde{\cal J}_\eps (\la,\eta,k)
= \frac{\eps^2}{2^5\pi^3} \int_{\bbR^3}d\beta_0d\beta_1d\beta_1'
\int_{\bbT^2}d\ell d\ell'  \bbE_\eps\left[
  \hat\psi(0,\ell) ^* \hat\psi(0,\ell')\right]
 \\
&
\times 
\frac{\tilde g(\eps\la/4+i\beta_1)}{[\eps\la/4-i(\beta_0+\beta_1)]
  [\eps\la/4+i(\beta_1 -\om(\ell))]
  [\eps\la/4+i(\beta_1-\om(k-\eps\eta/2))]}\\
&
\times 
\frac{\tilde g(\eps\la/4+i\beta_1')}{[\eps\la/4+i(\beta_0-\beta_1')]
  [\eps\la/4+i(\beta_1'+\om(\ell'))]
  [\eps\la/4+i(\beta_1'+\om(k+\eps\eta/2))]}.
\end{split}
\end{equation}
We integrate $\beta_1$ and $\beta_1'$ variables using the Cauchy integral formula
\begin{equation}
\label{CF}
\frac{1}{2\pi}\int_{\bbR}\frac{
  f(i\beta)d\beta}{z-i\beta}= f(z),\quad z\in \mathbb H,
\end{equation}
valid for any holomorphic  function $f$ on the right half-plane $\mathbb H:=[z\in\mathbb
C:\,{\rm Re}\,z>0]$ that belongs to the Hardy class $H^p(\mathbb H)$
for some $p\ge 1$,
see e.g. \cite[p. 113]{koosis}. Performing the above integration and, subsequently, changing variables $\eps\beta_0':=\beta_0+\om(k-\eps\eta/2)$ we get
\begin{equation}
\label{eq:tJ2b}
  \begin{split} 
&\tilde{\cal J}_\eps (\la,\eta,k)
= \frac{1}{2^3\pi\eps} \int_{\bbR}\frac{d\beta_0}{  \la/2-i \beta_0}
\int_{\bbT^2}d\ell d\ell'  \bbE_\eps\left[
  \hat\psi(0,\ell) ^* \hat\psi(0,\ell')\right]
 \\
&
\times 
\frac{|\tilde g\big(\eps\la/2-i\eps\beta_0+i \om(k-\eps\eta/2)\big)|^2}{
  \la/2-i \eps^{-1} (\om(\ell)-\om(k-\eps\eta/2) )-i \beta_0
  }\\
&
\times\frac{1}{
  \la/2+i \eps^{-1}  (\om(\ell') -\om(k-\eps\eta/2)  ) +i\beta_0}\\
&
\times 
\frac{1}{
  \la/2+i \eps^{-1} (\om(k+\eps\eta/2) -\om(k-\eps\eta/2)) +i\beta_0}.
\end{split}
\end{equation}
Change variables $\ell,\ell'$ according to the formulas
$\ell:=k'-\eps\eta'/2$ and $\ell':=k'+\eps\eta'/2$ and use (cf \eqref{nu}) 
$$
|\tilde g\big(\eps\la/2-i\eps\beta_0+i \om(k-\eps\eta/2)\big)|^2\approx |\nu(k)|^2,\quad
\mbox{as }\eps\ll1.
$$
We obtain then
\begin{equation}
\label{eq:tJ2b}
  \begin{split} 
&\tilde{\cal J}_\eps (\la,\eta,k)
\approx \frac{|\nu(k)|^2}{2^3\pi\eps} \int_{\bbR}\frac{d\beta_0}{  \la/2-i \beta_0}
\int_{ \bbR\times \bbT}\widehat{W}(0,\eta',k')d\eta' dk'  
 \\
&
\times 
\frac{1}{
  \la/2-i \eps^{-1} (\om(k'-\eps\eta'/2)-\om(k-\eps\eta/2) )-i \beta_0
 }\\
&
\times\frac{1}{
  \la/2+i \eps^{-1}  (\om(k'+\eps\eta'/2) -\om(k-\eps\eta/2)  ) +i\beta_0}\\
&
\times 
\frac{1}{
  \la/2+i \eps^{-1} (\om(k+\eps\eta/2) -\om(k-\eps\eta/2)) +i\beta_0}.
\end{split}
\end{equation}
Since $\om(k)$ is unimodal we can write
\begin{equation}
\label{eq:tJ2bd}
  \begin{split} 
&\tilde{\cal J}_\eps (\la,\eta,k)
\approx \frac{|\nu(k)|^2}{2^3\pi\eps} \sum_{\iota=\pm}\int_{\bbR}\frac{d\beta_0}{  \la/2-i \beta_0}
\int_{\bbR\times  [\iota k-\delta, \iota k+\delta]}\widehat{W}(0,\eta',k') d\eta'dk'
 \\
&
\times 
\frac{1}{
  \la/2-i \eps^{-1} (\om(k'-\eps\eta'/2)-\om(k-\eps\eta/2) )-i \beta_0
 }\\
&
\times\frac{1}{
  \la/2+i \eps^{-1}  (\om(k'+\eps\eta'/2) -\om(k-\eps\eta/2)  ) +i\beta_0}\\
&
\times 
\frac{1}{
  \la/2+i \eps^{-1} (\om(k+\eps\eta/2) -\om(k-\eps\eta/2)) +i\beta_0}.
\end{split}
\end{equation}
for a  (small) fixed $\delta>0$. Changing variables $k'=k+\eps\eta''$ and
using the approximations $\eps^{-1}[\om(k+\eps \xi)-\om(k)]\approx
\om'(k)\xi$ and $\widehat{W}(0,\eta',\iota k+\eps \eta'') \approx \widehat{W}(0,\eta',\iota k)$ we 
conclude that
\begin{equation}
\label{eq:tJ3}
  \begin{split} 
&\tilde{\cal J}_\eps (\la,\eta,k)
\approx \frac{|\nu(k)|^2}{2^3\pi} \sum_{\iota=\pm}\int_{\bbR}\frac{d\beta_0}{  \la/2-i \beta_0}
\int_{\bbR^2}\widehat{W}(0,\eta',\iota k+\eps \eta'') d\eta'd\eta''
 \\
&
\times 
\frac{1}{
  \la/2-i \eps^{-1} (\om(\iota (k+\eps\eta'')-\eps\eta'/2)-\om(k-\eps\eta/2) )-i \beta_0
 }\\
&
\times\frac{1}{
  \la/2+i \eps^{-1}  (\om(\iota (k+\eps\eta'')+\eps\eta'/2) -\om(k-\eps\eta/2)  ) +i\beta_0}\\
&
\times 
\frac{1}{
  \la/2+i \eps^{-1} (\om(k+\eps\eta/2) -\om(k-\eps\eta/2)) +i\beta_0}
\end{split}
\end{equation}
\begin{equation*}
  \begin{split} 
&
\approx \frac{|\nu(k)|^2}{2^3\pi} \sum_{\iota=\pm}\int_{\bbR}\frac{d\beta_0}{  (\la/2-i \beta_0) 
  (\la/2+i \om'(k)\eta +i\beta_0)}
\int_{\bbR}\widehat{W}(0,\eta',\iota k) d\eta'
 \\
&
\times \int_{\bbR}
\frac{1}{
  \la/2-i \om'(k)(\eta''+\eta/2-\iota\eta'/2)-i \beta_0
 }\\
&
\times\frac{d\eta''}{
  \la/2+i \om'(k)(\eta''+\eta/2+\iota\eta'/2) +i\beta_0}.
\end{split}
\end{equation*}
Integrating, first with respect to $\eta''$ and then $\beta_0$
variables, using e.g. \eqref{CF}, we get
\begin{equation}
\label{012208-20}
\lim_{\eps\to0+}\tilde{\cal J}_\eps (\la,\eta,k)=\frac{|\nu(k)|^2}{4|\bar\om'(k)|} \sum_{\iota=\pm}\int_{\bbR}\frac{d\eta}{  \la+i \om'(k)\eta }
\int_{\bbR}
\frac{\widehat{W}(0,\eta',\iota k) d\eta'}{
  \la+i\iota \om'(k)\eta'}
\end{equation}
From the second equality of \eqref{eq:tk01} we can see that formula
for $\tilde{\cal R}_\eps (\la,\eta,k)$ can be obtained from
\eqref{eq:tJ2b} by changing $\om(\ell)$ and $\om(\ell')$ to
$-\om(\ell)$ and $-\om(\ell')$ respectively and altering the complex
conjugation by the wave functions. It yields 
\begin{equation}
\label{eq:tJ2br}
  \begin{split} 
&\tilde{\cal R}_\eps (\la,\eta,k)
\approx \frac{|\nu(k)|^2}{2^3\pi\eps} \sum_{\iota=\pm}\int_{\bbR}\frac{d\beta_0}{  \la/2-i \beta_0}
\int_{\bbR\times  [\iota k-\delta, \iota k+\delta]}\widehat{W}(0,\eta',k') d\eta'dk'
 \\
&
\times 
\frac{1}{
  \la/2+i \eps^{-1} (\om(k'-\eps\eta'/2)+\om(k-\eps\eta/2) )-i \beta_0
 }\\
&
\times\frac{1}{
  \la/2-i \eps^{-1}  (\om(k'+\eps\eta'/2) +\om(k-\eps\eta/2)  ) +i\beta_0}\\
&
\times 
\frac{1}{
  \la/2+i \eps^{-1} (\om(k+\eps\eta/2) -\om(k-\eps\eta/2))
  +i\beta_0}\approx 0,
\end{split}
\end{equation}
as both the second and third lines are of order $\eps$, while the
fourth one is of order $1$. 

Summarizing, we have shown that (see \cite{kors18} for
 a rigorous derivation)
\begin{equation}
  \label{eq:15}
  \begin{split}
&\frac{1}{2\pi} \lim_{\eps\to0+}\int_{\bbR}e^{i\eta y} \widehat{
    W}_\varepsilon^2(t,\eta,k) d\eta\\
&
=\frac{ \gamma^2|\nu(k)|^2 }{4|\bar\omega'(k)|^2}1_{[[0, \bar\omega'(k)t]]}(y)
    \left( W(0,y- \bar\omega'(k) t,k) +
     W(0,-y+ \bar\omega'(k) t,-k)\right)
\end{split}
\end{equation}
and \eqref{eq:14b} follows.

\end{document}